\documentclass[twocolumn,superscriptaddress]{aastex63}

\accepted{to ACS Earth Space Chem on Sept 15, 2022}

\usepackage{graphicx}
\usepackage{bm}
\usepackage{mhchem}
\usepackage{longtable}
\usepackage{supertabular}

\usepackage{textpos}
\usepackage{hyperref} 
\usepackage{amsmath}
\usepackage{wasysym}

\setcitestyle{numbers,super}

\newcommand{\beginappendixA}{%
        \setcounter{table}{0}
        \renewcommand{\thetable}{S\arabic{table}}%
        \setcounter{figure}{0}
        \renewcommand{\thefigure}{S\arabic{figure}}%
        \setcounter{equation}{0}
        \renewcommand{\theequation}{S\arabic{equation}}%
     }

\definecolor{orang}{HTML}{E66100}
\definecolor{purps}{HTML}{5D3A9B}

\shorttitle{HCN production in the Hadean Earth atmosphere}
\shortauthors{Pearce et al.}

\begin{document}

\title{An experimental and theoretical investigation of HCN production in the Hadean Earth atmosphere}

\author{Ben K. D. Pearce*}
\affiliation{Department of Earth and Planetary Science, Johns Hopkins University, Baltimore, MD, 21218, USA}
\thanks{Corresponding author: bpearce6@jhu.edu}

\author{Chao He}
\affiliation{Department of Earth and Planetary Science, Johns Hopkins University, Baltimore, MD, 21218, USA}

\author{Sarah M. H{\"o}rst}
\affiliation{Department of Earth and Planetary Science, Johns Hopkins University, Baltimore, MD, 21218, USA}

\begin{abstract}
{\bf 
A critical early stage for the origin of life on Earth may have involved the production of hydrogen cyanide (HCN) in a reducing, predominantly \ce{H2} atmosphere. HCN is crucial for the origin of life as it is a possible precursor to several biomolecules that make up RNA and proteins including nucleobases, nucleotides, amino acids, and ribose. In this work, we perform an in depth experimental and theoretical investigation of HCN production in reducing atmospheric conditions (89--95\% \ce{H2}) possibly representing the earliest stages of the Hadean eon, $\sim$4.5--4.3 billion years ago. We make use of cold plasma discharges---a laboratory analog to shortwave UV radiation---to simulate HCN production in the upper layers of the atmosphere for \ce{CH4} abundances ranging from 0.1--6.5\%. We then combine experimental mass spectrum measurements with our theoretical plasma models to estimate the HCN concentrations produced in our experiments. We find that upper atmospheric HCN production scales linearly with \ce{CH4} abundance with the relation [HCN] = 0.13 $\pm$ 0.01[\ce{CH4}]. Concentrations of HCN near the surface of the Hadean Earth are expected to be about 2--3 orders of magnitude lower. The addition of 1\% water to our experiments results in a $\sim$50\% reduction in HCN production. We find that four reactions are primarily responsible for HCN production in our experiments: (i) \ce{^4N + CH3 -> H2CN + H -> HCN + H2}, (ii) \ce{^4N + CH -> CN + H} followed by \ce{CN + CH4 -> HCN + CH3}, (iii) \ce{C2H4 + ^4N -> HCN + CH3}, and (iv) \ce{^4N + ^3CH2 -> HCN + H}. The most prebiotically favorable Hadean atmosphere would have been very rich in \ce{CH4} ($>$ 5\%), and as a result of greenhouse effects the surface would be likely very hot. In such a prebiotic scenario, it may have been important to incorporate HCN into organic hazes that could later release biomolecules and precursors into the first ponds.
}
\end{abstract} 

\keywords{early Earth --- astrobiology --- atmospheric chemistry --- origin of life --- HCN production}

\section{Introduction}

The origin of life on Earth occurred in the Hadean eon, roughly between 4.5 and 3.7 billion years ago (bya) (e.g., \citet{2018AsBio..18..343P,2002Natur.418..214J}). At this time, comets and asteroids were impacting the surface of the Earth at a rate 100 thousand to 100 million times greater than today \citep{Ferus_et_al2021,1990Natur.343..129C}. It is expected that a large fraction of the impactors were similar in composition to enstatite meteorites, and contained a significant amount of iron \citep{CatlingZahnle2020}. This is based on the similarities of isotopes and relative proportions of lithophile (O, Ca, Ti, and Nd), moderately siderophile (Cr, Ni, and Mo), and highly siderophile (Ru) elements in the Earth's mantle compared to enstatite meteorites \citep{Dauphis_2017,CatlingZahnle2020}. Furthermore, zircon analyses suggest the Hadean Earth had surface water by 4.4 bya \citep{2001Natur.409..178M,2001GeCoA..65.4215P}, indicating that warm little pond environments were present for the emergence of life \citep{Damer_Deamer2019}. The oxidation of this iron by water (i.e. \ce{Fe + H2O -> FeO + H2}) released large amounts of \ce{H2} into the atmosphere (e.g. 60--90\% \citep{Pearce_et_al2022a,Zahnle_et_al2020}), providing reducing conditions favorable for the production of hydrogen cyanide (HCN): a key biomolecule precursor. As the impact rate steadily decreased from 4.5 to 3.7 bya, the hydrogen content of the atmosphere dropped as a result of hydrodynamic escape from the upper atmosphere. By $\sim$4.3 bya, the bombardment rate was likely too low to maintain \ce{H2}-dominant conditions \citep{Pearce_et_al2022a}, and an atmosphere dominated by \ce{N2}/\ce{CO2} from outgassing would follow \citep{Reference119,Levine1985}. HCN, and its precursor \ce{CH4} are rapidly destroyed in oxidizing (\ce{CO2}-rich) environments, e.g., via the OH radical \citep{Pearce_et_al2022a}.



The famous Miller-Urey experiment was the first demonstration of abiotic biomolecule production in simulated \ce{H2}-dominated early Earth conditions \citep{Miller1953}. Using electrodes attached to a glass flask filled with reducing gases (\ce{H2}, \ce{CH4}, \ce{NH3} and \ce{H2O}), Miller simulated lightning strikes to produce atmospheric HCN \citep{Miller_Schlesinger1983,Reference440}. Attached below the flask was a condenser, where this HCN (along with other water soluble species) rained out into a reservoir and reacted in aqueous solution to produce biomolecules. Many have repeated this famous experiment to demonstrate the production of dozens of amino acids as well as the four canonical RNA nucleobases in atmospheres containing various reducing and oxidizing gas mixtures (e.g., \citet{Ring_et_al1972,Wolman_et_al1972,Miller_Schlesinger1983,Schlesinger_Miller1983,Miyakawa_et_al2002b,Cleaves_et_al2008,Reference438}).

In general, reduced carbon is required as a precursor to produce atmospheric HCN. In contrast, the pathway to HCN from oxidized carbon (i.e., \ce{CO2} and \ce{CO}) is unfavorable, as there is a large energetic barrier to remove oxygen from these species. \ce{CH4} is considered the dominant source of reduced carbon in planetary atmospheres, as it is typically the most abundant compared to other reduced carbon species (e.g., \ce{C2H6}, \ce{C2H4}). Equilibrium models suggest that early hydrothermal systems on Earth could produce up to $\sim$2.5 ppm of atmospheric \ce{CH4} \citep{Guzman-Marmolejo2013}. On the other hand, very large impactors similar in size to Vesta ($\sim$400+ km) have been proposed as a source of $\sim$0.1--10\% atmospheric \ce{CH4} \citep{Zahnle_et_al2020}. Abundant \ce{CH4} is unlikely to have come from volcanism \citep{Wogan_et_al2020}; However, molecular nitrogen would have been present as a result of volcanic outgassing \citep{Reference119,Levine1985}, supplying nitrogen atoms for HCN.

Several atmospheric photochemical models have identified a link between HCN production and atmospheric \ce{CH4} abundance (e.g., \citet{Pearce_et_al2022a,Zahnle_et_al2020,2019Icar..329..124R,2011EPSL.308..417T,Reference591}). However, a relation between these two species based on models and experiments has not been calculated.

\citet{Miller_Schlesinger1983} are the only ones that have performed spark-discharge experiments to investigate the effects of \ce{CH4} abundance on HCN production in reducing atmospheres. For \ce{CH4} abundances ranging from 17--50\%, they found relatively constant $\sim$10\% yields of HCN. However, such high atmospheric \ce{CH4} concentrations would be unlikely on early Earth, even in typical large impact scenarios \citep{Zahnle_et_al2020}. 

Considering the importance of HCN as a precursor for biomolecule production, it is of great interest to understand how HCN production scales with \ce{CH4} abundance for plausible Hadean Earth atmospheres. A novel and comprehensive approach to exploring this relation is to couple laboratory experiments with theoretical chemical kinetics simulations. The advantages to this approach are: 1) ground-truth experiments provide a means to validate theory, 2) theoretical simulations provide a way to estimate the HCN concentrations produced in experiments, without the difficulty and danger of having to run pure HCN gas standards, and 3) theoretical simulations provide a pathway to discover which reactions are responsible for HCN production in the studied atmospheres.


Atmospheric HCN production requires an energy source to dissociate \ce{N2} and \ce{CH4} into reactive radicals. For example, \ce{^4N + CH3 -> HCN + H2} is one of the dominant reactions for HCN production in Titan's atmosphere (e.g., \citet{Pearce2020a}). Recent models suggest that UV radiation was the main energy source for HCN production in the Hadean atmosphere. In contrast, lightning's contribution to the global HCN concentration was negligible---even when considering volcanic electrical storms \citep{Pearce_et_al2022a}. 
Photodissociation of \ce{N2} requires very short UV wavelengths ($\lesssim$ 110 nm) that are outside the spectrum of most UV lamps \citep{ResonanceLtd}. For this reason, cold plasma discharges (i.e., electron/ion collisions) can be used as a laboratory analog to excite and dissociate \ce{N2} similarly to short wavelength UV radiation (e.g., \citet{Cable_et_al2012}). We note that HCN and biomolecules can also be produced in high-velocity impacts \citep{Ferus_et_al2020}, and delivered to early Earth via comets \citep{Todd_Oberg2020,2016SciA....2E0285A}.

In this paper, we perform laboratory experiments to simulate the production of atmospheric HCN as function of \ce{CH4} abundance on the Hadean Earth. We make use of the Planetary Haze Research (PHAZER) experimental setup, which allows us to apply cold plasma discharges to different gas mixtures and detect changes in mass spectra resulting from HCN production with a Residual Gas Analyzer (RGA). We then perform 0D chemical kinetic calculations using three different chemical networks and compare the resulting theoretical HCN concentrations with the increases to the experimental 26 peak from the plasma experiments. From this we are able to calculate a relation, based on experiment and theory, between HCN production and \ce{CH4} abundance for our reducing atmospheric experiments. This relation can also be applied to theoretical HCN production via shortwave UV radiation in the upper atmospheric layers of the Hadean Earth. Lastly, we perform sensitivity analyses on the theoretical simulations to determine which reactions are responsible for HCN production and destruction for each simulated experiment.


\section{Methods}

\subsection{PHAZER Experiments}

\begin{figure*}[!hbtp]
\centering
\includegraphics[width=\linewidth]{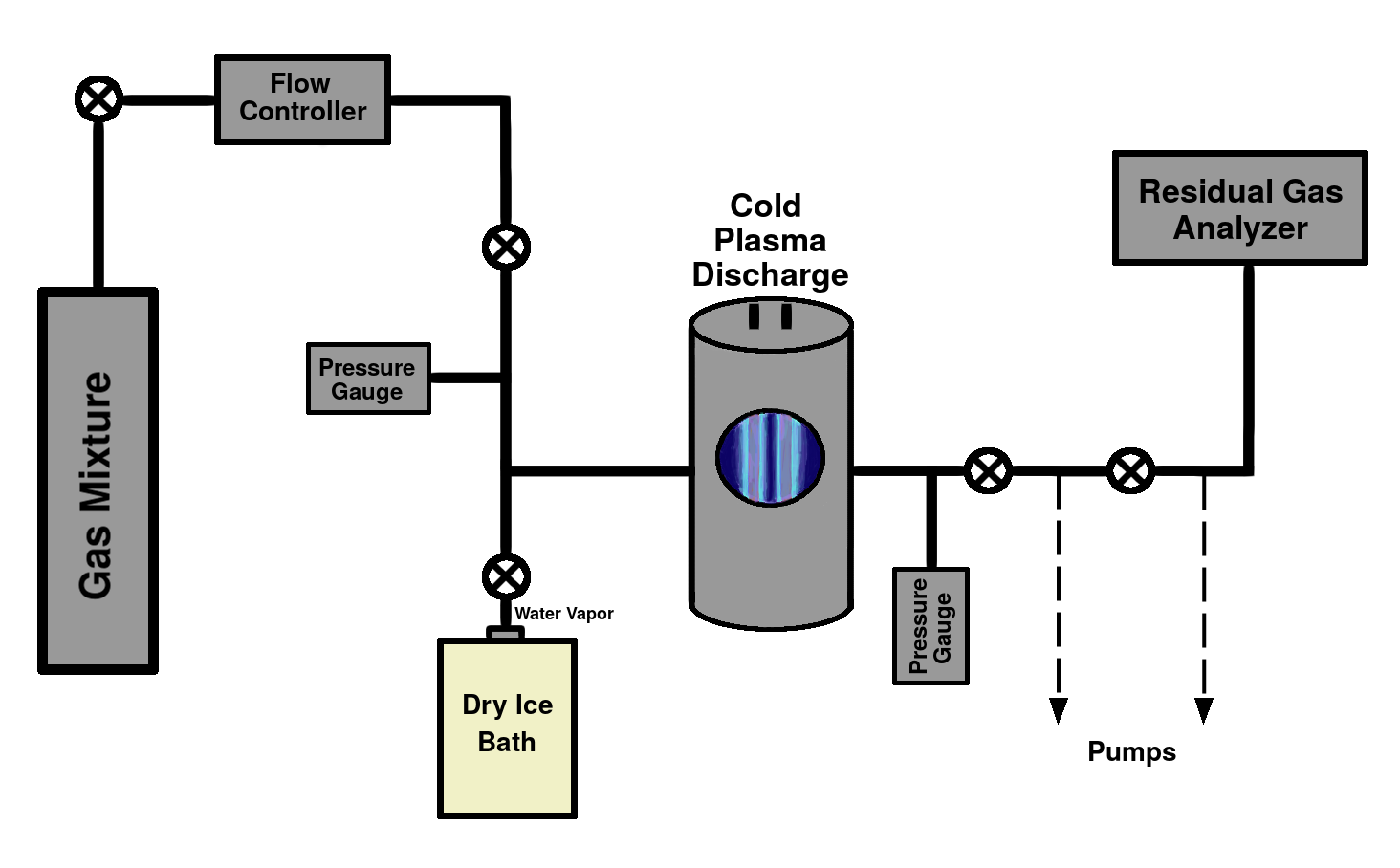}
\caption{A schematic of the PHAZER apparatus used for our Hadean atmospheric experiments.}
\label{PHAZER}
\end{figure*}

In Figure~\ref{PHAZER}, we display a schematic of the PHAZER experimental setup \citep{He_et_al2017} used for our Hadean Earth experiments. PHAZER is a flow system with a stainless steel chamber that is used to simulate the chemistry occurring in planetary atmospheres. Within the chamber is the option to attach two electrodes that produce a cold plasma discharge when a voltage differential is applied. The electrons and ions produced by the AC glow discharge are roughly in the 1.5 to 18.5 eV range, and are energetic enough to directly dissociate all four of our input species (\ce{H2}, \ce{H2O}, \ce{CH4}, and \ce{N2}). The total gas temperature is not significantly altered by the plasma, and remains at roughly room temperature. Although the Hadean upper atmosphere was likely somewhat cooler than this, we do not adjust the chamber temperature for simplicity. When decreasing the temperature by 50$^{\circ}$C in our CRAHCN-O chemical kinetic model of Experiment 1, we see only a $\sim$14 \% decrease in HCN production. Such a setup has been used in the past as a laboratory analog for planetary UV and lightning chemistry (e.g., \citet{2012AsBio..12..809H,Ferus_et_al2017b,Reference438}). The molecules produced in the cold plasma discharge flow via stainless steel piping to the RGA quadrupole mass spectrometer for identification and measurement.

Our experimental procedure is similar to previous PHAZER studies \citep{He_et_al2018a,He_et_al2018b,Horst_et_al2018a,He_et_al2019,He_et_al2020a,He_et_al2020b,Moran_et_al2020}. First, we prepare the initial gas mixtures using high-purity gases (\ce{H2}-99.9999\%, \ce{N2}-99.9997\%, \ce{CH4}-99.999\%; Airgas) and high performance liquid chromatography-grade water (Fisher Chemical). For the gas mixture with $\sim$1\% water, the partial pressure of vapor is roughly controlled by putting a cylinder of water in a dry ice-methanol-water cooling bath maintained in the -42 to -46$^{\circ}$C range. We wait at least 2 hours for the water temperature to equilibrate with the cooling bath before turning on the gas flow.

We obtain two background spectra (30-scan averages) of the RGA chamber prior to each experiment and subtract the average of these spectra from the mass spectra measurements of the gas mixtures during each experiment. The RGA is set to the standard 70 eV ionization energy with a scanning mass range of 1--100 AMU.

To begin each experiment, we flow the gas mixture continuously through the system at 5 standard cubic meters per minute. We then obtain two histogram spectra and one analog spectrum of the input gas mixtures prior to turning on the plasma source. All spectra measurements are an average of 30 consecutive scans. We wait 20 minutes after turning on the 170 W/m$^2$ plasma source before measuring the spectra with the RGA in order to ensure the experimental composition reaches a steady state. Then, we obtain 8--9 histogram spectra and one analog spectra with the AC glow discharge turned on, over the course of 4--5 hours.

In Table~\ref{ExperimentCompositions}, we display the gas compositions for our five Hadean Earth experiments. These compositions are \ce{H2}-dominant - in line with the expected reducing conditions produced by the oxidation of iron from enstatite impactors at $\sim$4.5--4.3 bya \citep{Pearce_et_al2022a,CatlingZahnle2020,Zahnle_et_al2020}. We maintain the nitrogen content at 5\% for all experiments, which is expected to be present from volcanic outgassing \citep{Reference119,Levine1985}. We also include water at 1\% in one experiment to see whether a large atmospheric water abundance affects our results. We note that no unexpected or unusually high safety hazards were encountered in performing these experiments.

\begin{table}[ht!]
\centering
\caption{Input molar compositions (in \%) for six Hadean Earth experiments. Each dry experiment is run twice and Experiment 6 is run once with water to see how it affects HCN production.\label{ExperimentCompositions}} 
\begin{tabular}{lcccccc}
\\
\multicolumn{1}{c}{Gas} &  
\multicolumn{1}{l}{Exp. 1} & 
\multicolumn{1}{l}{Exp. 2} &
\multicolumn{1}{l}{Exp. 3} & 
\multicolumn{1}{l}{Exp. 4} & 
\multicolumn{1}{l}{Exp. 5} &
\multicolumn{1}{l}{Exp. 6}    
\\[+2mm] \hline \\[-2mm]
\ce{CH4} & 6.5 & 3.6 & 1.5 & 0.5 & 0.1 & 3.6\\
\ce{N2} & 5 & 5 & 5 & 5 & 5 & 5\\
\ce{H2} & 89 & 91 & 94 & 95 & 95 & 90\\
\ce{H2O} & 0 & 0 & 0 & 0 & 0 & 1\\
\\[-2mm] \hline
\end{tabular}
\end{table}


\subsection{Chemical Kinetics Models}

We run 0D closed system chemical kinetic models at room temperature to simulate our experimental conditions using the KINTECUS software package \citep{kintecus19}. We use three different chemical networks that are commonly used to determine HCN chemistry as input into our models to see if the results are network dependent. These networks are: 1) CRAHCN-O \citep{Reference598,Pearce2020a,Pearce2020b,Pearce_et_al2022a}, 2) H{\'e}brard12 \citep{2012AA...541A..21H}, and 3) Venot15 \citep{Venot2015}. We add 53 new reactions to CRAHCN-O in order to better estimate the production of hydrocarbons and other potential species of interest such as \ce{NO} and \ce{NH3}. We also remove 12 reactions from the previous versions of CRAHCN-O that produced species that had no sink reactions in our network and led to erroneous build up of these species (e.g., \ce{OH + H2CN + M -> H2CNOH + M} and \ce{OH + NH + M -> HNOH + M}). The new reactions added to CRAHCN-O are displayed in Table S1 in the Supporting Information. In Table~\ref{ChemicalNetworks}, we summarize the three chemical networks.

\begin{table*}[ht!]
\centering
\caption{Summary of the three chemical networks input in our chemical kinetic models. \label{ChemicalNetworks}} 
\begin{tabular}{lcccc}
\\
\multicolumn{1}{c}{Network} &  
\multicolumn{1}{l}{Species} & 
\multicolumn{1}{l}{Gas Phase Reactions} &
\multicolumn{1}{l}{Plasma Dissociation Reactions} &
\multicolumn{1}{c}{Source} 
\\[+2mm] \hline \\[-2mm]
CRAHCN-O & 49 & 300 & 17 & \citet{Pearce2020a,Pearce2020b,Pearce_et_al2022a} + this work \\
H{\'e}brard12 & 136 & 788 & 16 & \citet{2012AA...541A..21H} + this work \\
Venot15 & 238 & 2002 & 16 &  \citet{Venot2015} + this work\\
\\[-2mm] \hline
\end{tabular}
\end{table*}

CRAHCN-O was developed to model HCN and \ce{H2CO} chemistry in planetary atmospheres dominated by any of \ce{H2}, \ce{N2}, \ce{CH4}, \ce{CO2}, and \ce{H2O}. It is a reduced network containing experimental rate coefficients when available, and consistently calculated quantum chemical rate coefficients otherwise \citep{Pearce2020a,Pearce2020b,Pearce_et_al2022a}. CRAHCN-O has been used to model HCN production in Titan's atmosphere and produces results that agree with the observations of HCN taken by the Cassini spacecraft \citep{Pearce2020a}. CRAHCN-O has also been used to model HCN and \ce{H2CO} production in the Hadean Earth atmosphere \citep{Pearce_et_al2022a}.

The H{\'e}brard12 network was developed to model HCN and HNC production in Titan's atmosphere \citep{2012AA...541A..21H}, and also produces results that agree with the observational measurements of HCN by the Cassini spacecraft. The H{\'e}brard12 network contains experimental, theoretical, and estimated rate coefficients for 788 reactions, and is more complete than CRAHCN-O for species with carbon numbers 4 $\leq$ C $\leq$ 8. 

The Venot15 network was developed to study the chemistry of warm carbon-rich exoplanet atmospheres \citep{Venot2015}. It is the largest of the networks used in this study, containing experimental, theoretical, and estimated rate coefficients for 2002 gas phase reactions. The Venot15 network is complete for carbon species up to C $\leq$ 14.


Electron impact dissociation rate coefficients from our plasma source take the form

\begin{equation}
    \ce{A + e^- -> B + C + e^-}
\end{equation}\label{Equation1}

We use the following standard rate coefficient equation for electron-induced processes \citep{Alves_et_al2018} to estimate the dissociation rate coefficients for each of the species for which electron impact dissociation data was available.

\begin{equation}
k = \sqrt{2q/m_e} \int_{1.5eV}^{18.5eV} \epsilon^{1/2} \sigma(\epsilon) f(\epsilon) d\epsilon 
\end{equation}
where $k$ is the dissociation rate coefficient for a particular electron energy in m$^3$ s$^{-1}$, $q$ is used to convert J to eV (1.602$\times$10$^{-19}$ J/eV), $m_e$ is the mass of an electron (9.11$\times$10$^{-31}$ kg), $\epsilon$ is the electron energy (eV), $\sigma(\epsilon)$ is the electron impact dissociation cross-section as a function of electron energy in m$^{-2}$, and $f(\epsilon)$ is the electron energy distribution function (EEDF) (eV$^{-1}$).

We run simulations using both Maxwellian and Druyvesteyn EEDFs to understand the importance of EEDF choice on our chemical kinetic results. If the electrons are in thermodynamic equilibrium with the surrounding plasma, their distribution function is expected to be Maxwellian. Maxwellian EEDFs are conventionally used in plasma theory; however, low-pressure discharges such as those in our experiments are generally non-Maxwellian. At a pressure of 0.3 Torr, one study found ionized plasmas to be Druyvesteyn-like \citep{Godyak_et_al1993}. This is within a factor of 6 of our typical chamber pressure ($\sim$1.8 Torr). These two EEDFs take the form \citep{Gudmundsson2001,khalilpour_foroutan_2020}:

\begin{equation}
f(\epsilon) = c_1 \epsilon^{1/2} e^{- c_2 \epsilon^x}
\end{equation}
where x = 1 for a Maxwellian distribution, and x = 2 for a Druyvesteyn distribution.

Coefficients $c_1$ and $c_2$ are defined as:
\begin{equation}
c_1 = \frac{x}{<\epsilon>^{3/2}} \frac{\Gamma(\frac{5}{2x})^{3/2}}{\Gamma(\frac{3}{2x})^{5/2}}
\end{equation}
\begin{equation}
c_2 = \frac{1}{<\epsilon>^{x}} \frac{\Gamma(\frac{5}{2x})^{x}}{\Gamma(\frac{3}{2x})^{x}}
\end{equation}
where $<\epsilon>$ is the average electron energy (eV), and $\Gamma$ is the Gamma function.

\begin{figure}[!hbtp]
\centering
\includegraphics[width=\linewidth]{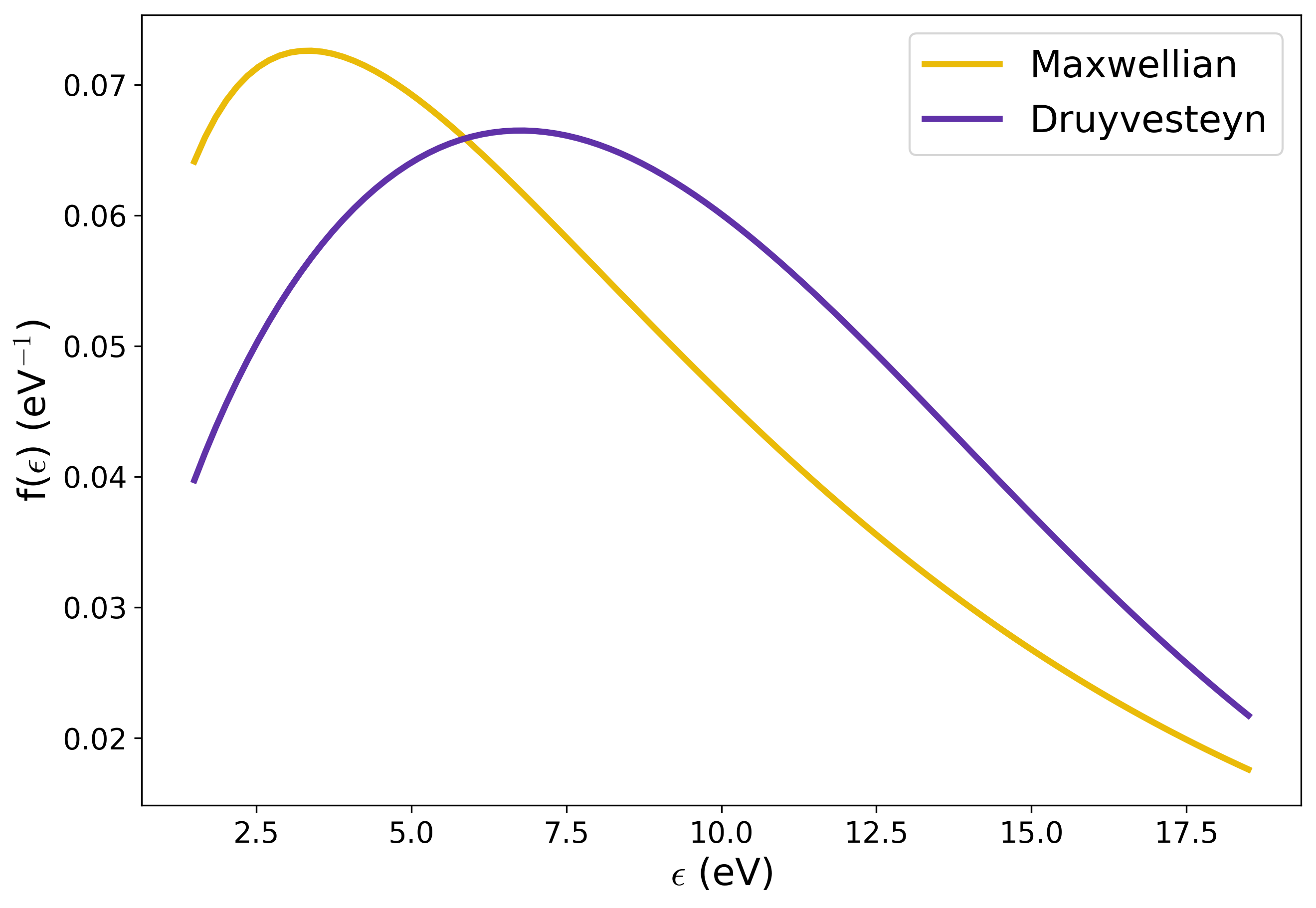}
\caption{Analytical electron energy distribution functions (EEDFs) for the PHAZER plasma source for the two distributions explored in this work. \label{EEDFs}}
\end{figure}

In Figure~\ref{EEDFs}, we plot the analytical EEDFs, assuming an average electron energy in the center of the range for our plasma source, i.e. 10 eV. Studies of EEDFs from DC glow discharge in \ce{H2}-dominant atmospheres suggest the average negative glow region to be roughly 6 and 8 eV for \ce{H2} abundances of 70 and 80\%, respectively \citep{El-Brulsy_et_al2012}. Given our AC glow discharge experiments are roughly \ce{H2} = 90\%, 10 eV may be a reasonable choice for an average electron energy.

We reduce Equation~\ref{Equation1} to a first order rate coefficient (s$^{-1}$) by multiplying $k$ by the electron number density $n$ (m$^{-3}$). This is reasonable under the assumption that the plasma discharge maintains a constant density of electrons. We estimate the electron number density assuming that the plasma energy density (170 W m$^{-2}$) is primarily electrons, and the average electron energy $<\epsilon>$ = 10 eV. The electron number density is calculated using the equation below.

\begin{equation}
    n = \frac{\omega}{<\epsilon> q v_{th}}
\end{equation}
where $\omega$ is the plasma energy density (170 W m$^{-2}$), and $v_{th}$ is the most probable thermal velocity of a free electron ($v_{th}$ = $\sqrt{2 k_B T/m_e}$ = $\sqrt{4\epsilon_{th}/3m_e}$, where $\epsilon_{th}$ = $\frac{3}{2} k_B T$ is the mean thermal energy of the electrons in the plasma (10 eV).

In Table~\ref{DissociationReactions}, we display the calculated dissociation rate coefficients used in our study along with the sources for electron impact dissociation cross-sections $\sigma(\epsilon)$.

\begin{table*}[ht!]
\centering
\caption{The first-order electron impact dissociation rate coefficients calculated for our plasma source, for Maxwellian and Druyvesteyn EEDFs. The sources of cross-section data are also listed. For \ce{NO}, \ce{C2H6}, \ce{C2H4}, \ce{CN}, and \ce{CH3CN}, partial ionization (e.g., \ce{C2H6 + e^- -> C2H4^+ + H2 + 2 e^-}) or dissociative electron attachment (e.g., \ce{HCN + e^- -> H + CN^-}) cross sections are used to represent neutral dissociation cross sections. This is because our reaction networks only contain neutral species. \label{DissociationReactions}} 
\begin{tabular}{lccl}
\\
\multicolumn{1}{c}{Reaction} &  
\multicolumn{1}{l}{k$_{Maxwellian}$ (s$^{-1}$)} & 
\multicolumn{1}{l}{k$_{Druysteyn}$ (s$^{-1}$)} &
\multicolumn{1}{l}{Cross-section source}
\\[+2mm] \hline \\[-2mm]
\ce{N2 + e^- -> ^4N + ^4N + e^-} & 0.31 & 0.42 & \citet{Tabata_et_al2006} \\
\ce{H2 + e^- -> H + H + e^-} & 0.28 & 0.38 & \citet{Yoon_et_al2008} \\
\ce{CH4 + e^- -> CH3 + H + e^-} & 0.43 & 0.58 & \citet{Bouman_et_al2021} \\
\ce{CH4 + e^- -> ^3CH2 + H2 + e^-} & 0.20 & 0.27 & \citet{Bouman_et_al2021} \\
\ce{H2O + e^- -> OH + H + e^-} & 0.15 & 0.20 & \citet{Itikawa_Mason2005} \\
\ce{O2 + e^- -> ^3O + ^1O + e^-} & 0.072 & 0.096 & \citet{Cosby1993} \\
\ce{CO + e^- -> C + ^3O + e^-} & 0.18 & 0.24 & \citet{Shirai_et_al2001} \\
\ce{CO2 + e^- -> CO + ^1O + e^-} & 0.048 & 0.065 & \citet{Shirai_et_al2001} \\
\ce{OH + e^- -> ^3O + H + e^-} & 0.093 & 0.13 & \citet{Chakrabarti_et_al2019} \\
\ce{OH + e^- -> ^1O + H + e^-} & 0.044 & 0.06 & \citet{Chakrabarti_et_al2019} \\
\ce{NO + e^- -> ^4N + ^3O + e^-} & 1.8$\times$10$^{-3}$ & 2.2$\times$10$^{-3}$ & \citet{Song_et_al2019} \\
\ce{C2H4 + e^- -> C2H2 + H2 + e^-} & 0.14 & 0.19 & \citet{Shirai_et_al2002} \\
\ce{C2H6 + e^- -> C2H4 + H2 + e^-} & 0.42 & 0.56 & \citet{Shirai_et_al2002} \\
\ce{C2H6 + e^- -> CH3 + CH3 + e^-} & 0.017 & 0.022 & \citet{Shirai_et_al2002} \\
\ce{HCN + e^- -> CN + H + e^-} & 3.3$\times$10$^{-3}$ & 2.2$\times$10$^{-3}$ & \citet{May_et_al2010} \\
\ce{C2H2 + e^- -> C2H + H + e^-} & 0.19 & 0.30 & \citet{Song_et_al2017} \\
\ce{CH3CN + e^- -> CH3 + CN + e^-} & 1.5$\times$10$^{-5}$ & 1.7$\times$10$^{-5}$ & \citet{Sailer_et_al2003} \\
\\[-2mm] \hline
\end{tabular}
\end{table*}




\section{Results - Experiments}

\begin{figure*}[!hbtp]
\centering
\includegraphics[width=\linewidth]{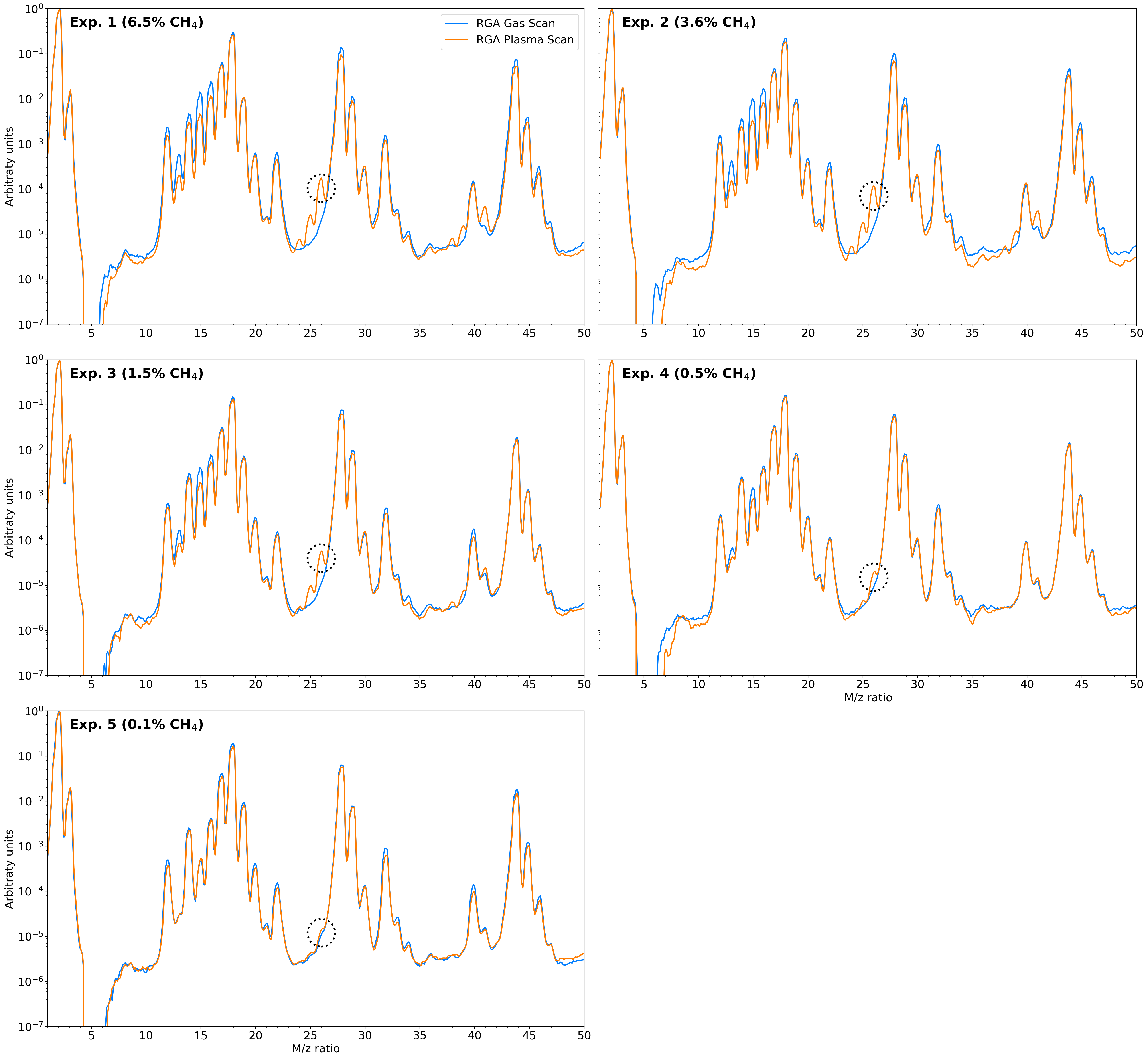}
\caption{Analog mass spectra for each of the five Hadean Earth experimental compositions, normalized to the 2 (\ce{H2}) peak. Included are scans of the input gas mixtures, as well as the compositions after cold plasma discharge has been turned on. Each mass spectrum is an average of 30 consecutive scans. The 26 peak is circled in each plot as this is the peak with which HCN can be detected in our experiments (via the \ce{CN^+} fragment). Conversely, any HCN in the 27 peak is completely washed out by the tail from the 28 peak of \ce{N2}. \label{Fig1Analog}}
\end{figure*}

In Figure~\ref{Fig1Analog}, we display the analog RGA mass spectra for Experiments 1--5. Analog plots are valuable for verifying the presence or absence of distinct peaks at each mass-to-charge ratio. Both the initial gas mixtures and the compositions after the plasma source is turned on are displayed. 

In all plots, we can see that the peaks corresponding to the mass spectrometry fragment pattern of \ce{CH4}, i.e., m/z = 16, 15, 14, 13, decrease after plasma is turned on, suggesting methane has been converted into other carbon species. We also see that the 26 peak, corresponding to HCN or hydrocarbons (e.g., \ce{C2H6}, \ce{C2H4}, \ce{C2H2}) increases after plasma is turned on, and that the increase correlates with the abundance of \ce{CH4} in the experiment. The 27 peak, which is the main detection fragment for HCN, is completely washed out by the tail of the intense 28 peak of \ce{N2} and \ce{CO}. Given the mass spectrometry fragmentation peak for HCN at m/z = 27 is ~5.9 times greater than the HCN fragmentation peak at m/z = 26 \citep{NISTChemWebBook_massspec}, we can only detect a HCN peak at m/z = 27 when the tail of the 28 peak from N2 and CO bleeding into m/z = 27 is $<$ 5.9 times greater than the HCN peak at m/z = 26. In Figure~\ref{Fig1Analog}, we see that the signal at m/z = 27 from the \ce{N2}/\ce{CO} tail is 7.2, 6.8, 10, 23, and 29 times higher than the signal at m/z = 26 for experiments 1--5, respectively.

However, the 26 peak alone cannot be used to determine HCN production, given that it can also represent hydrocarbon production. The increase in the peaks at 24 and 25 after plasma is turned on suggest that hydrocarbons (e.g., \ce{C2H2} and \ce{C2H4}) are produced in these experiments. For this reason, we run chemical kinetic models for each experiment to deduce the mostly likely source of the increase in 26 peaks from cold plasma discharge.

\begin{figure*}[!hbtp]
\centering
\includegraphics[width=\linewidth]{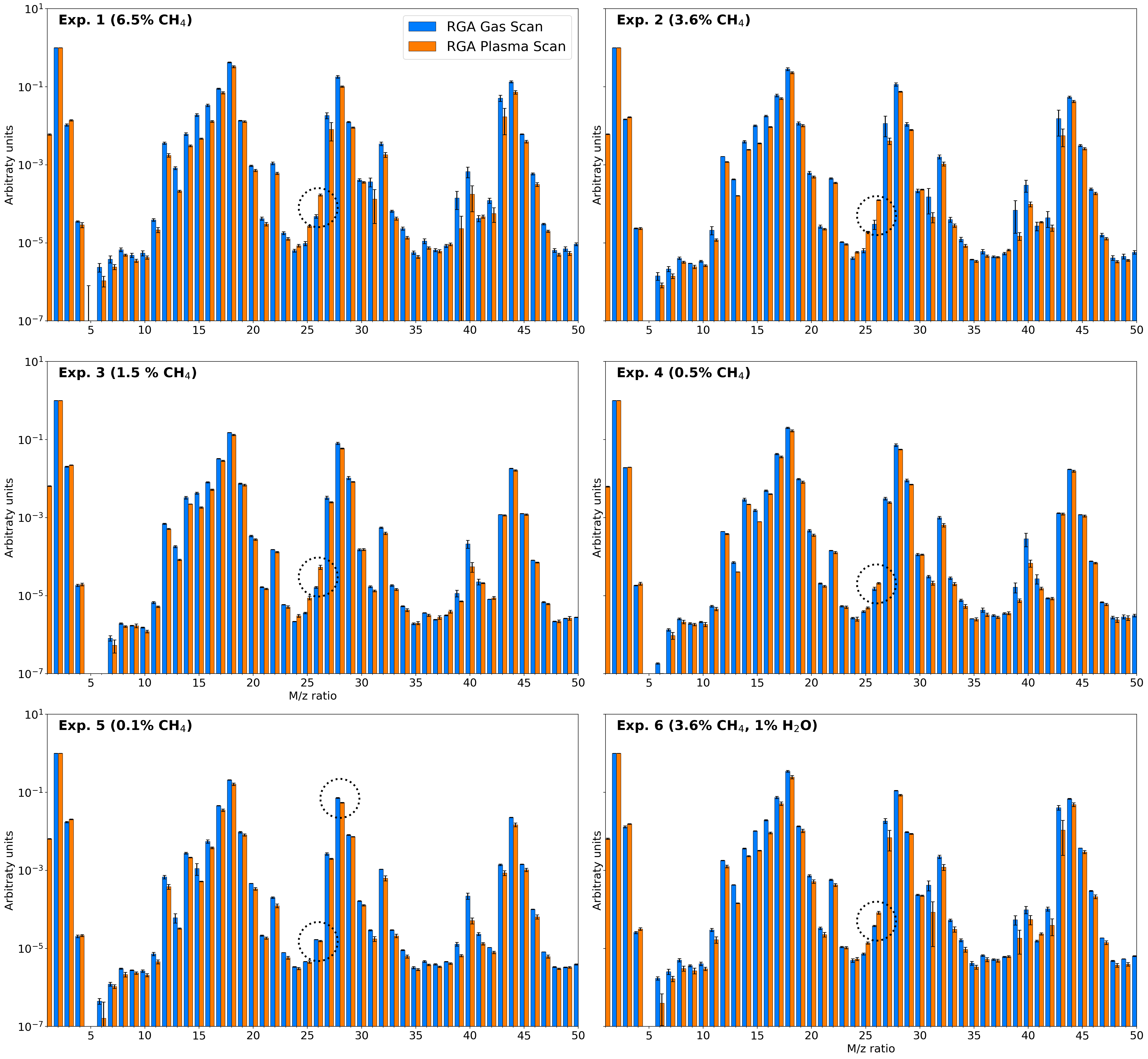}
\caption{Histogram mass spectra for Run 1 of Experiments 1--5, as well as a single Experiment 6 run with 1\% water vapor. Run 2 data can be found in the Supporting Information. The 26 peak is circled in each plot to direct the reader's eye to the location where HCN can be detected in our experiments. Experiment 5 also has peak 28 circled to emphasize that the 26/28 ratio increases after plasma is turned on. \label{histograms1}}
\end{figure*}

In Figures~\ref{histograms1} and S1, we plot the histogram RGA mass spectra of Runs 1 \& 2 of Experiments 1--5, and Experiment 6 with 1\% water. Histogram plots are valuable for performing quantitative analyses.

Similar to the analog plots, we see a clear increase in the 26 peaks after plasma is turned on. The only exception is for the weakest signal in Run 1 of Experiment 5, for which we can consider the 26/28 peak ratio to see the increase. This ratio is the more critical factor for HCN detection in histogram scans, because for small ``shoulder'' peak increases at m/z = 26 as we see for Experiment 5 in Figure~\ref{Fig1Analog}, the tail from the 28 peak makes up a significant portion of the signal. Therefore, if the 28 peak decreases after plasma is turned on, then the component of the signal at m/z = 26 that is due to the 28 tail decreases. Considering the 26/28 ratio accounts for this and allows us to measure the weaker signals of Experiment 5.

Because we know that the signals at m/z = 27 are completely due to the 28 tails, it is not surprising to see that the 27 peaks in Figures~\ref{histograms1} and S1 decrease after plasma is turned on, along with the decrease in the 28 peaks. 


We see increases to the 24 and 25 peaks, or at least the 24/28 and 25/28 ratios, due to the cold plasma discharge in all experiments. These signals are most likely due to hydrocarbon production. \ce{C2H2}, \ce{C2H4}, \ce{C2H6}, and \ce{C3H8} have fragment peaks for m/z = 24 and 25 of 0.05 and 0.19, 0.02 and 0.08, 0.005 and 0.035, 0.001 and 0.005, respectively \citep{NISTChemWebBook_massspec}, we might expect \ce{C2H2} and \ce{C2H4} to be the main contributors to these peaks. However, again, we look to our chemical kinetics models in the following section to deduce the most likely source of these signals.

The reason that coupling chemical kinetics to these experiments is so valuable, is that there are often multiple combinations of molecular species and concentrations that can produce identical signals in the mass spectra. Without chemical kinetics, it is not always possible to determine the correct combination. In the past, authors have demonstrated the use of Monte Carlo methods to obtain a statistical distribution of potential compositions \citep{Gautier_et_al2020}. Here, we demonstrate the use of plasma modeling with chemical kinetics in order to obtain a single most-likely solution. Experimentally, we find the increase in the 25 peak relative to the 26 peak (normalized to the 28 peak), to range from 0.08:1 to 0.37:1, with an average of (0.16$\pm$0.07):1. If we remove the 0.37:1 outlier, corresponding to Experiment 5 Run 1 which had the weakest 25 and 26 peaks, the average is (0.14$\pm$0.02):1. Given Experiment 5 (0.1\% \ce{CH4}) is at the detection limit for HCN production using our instruments, we suggest the removal of this outlier may provide a more reasonable estimate. Regardless, the following exercise to show that multiple solutions can produce this ratio works similarly whether the ratio is 0.14:1 or 0.16:1. We can estimate possible solutions for this signal with the equation:

\begin{equation}
    \frac{\sum_{m/z=25}{n_i \sigma_i f_i}}{\sum_{m/z=26}{n_i \sigma_i f_i}}
\end{equation}
where $n_i$ is the molar concentration of species $i$, $\sigma_i$ is the ionization cross-section for species $i$ at 70 eV, and $f_i$ is the fragmentation fraction for species $i$ at the specified m/z peak. The ionization cross-sections for \ce{HCN}, \ce{C2H2}, and \ce{C2H4} at 70 eV are 3.6 $\AA^{2}$, 4.4 $\AA^{2}$, and 5.1 $\AA^{2}$, respectively \citep{Ionization_Cros2004,Pandya_et_al2012}. The fragmentation fractions for theses species at m/z=25 and m/z=26 are 0 and 0.17, 0.19 and 1, and 0.08 and 0.53, respectively \citep{NISTChemWebBook_massspec}.

Three possible solutions for an increase in the 25 peak relative to the 26 peak that match our experimental average of 0.14 are (in molar percentages HCN/\ce{C2H2}/\ce{C2H4}): 60/20/20, 20/0/80, and 70/30/0. By modeling our experiments using chemical kinetics, we can calculate the expected concentrations for each of theses species and determine the most likely solution.

In the mass spectra of Experiment 6 with water in Figure~\ref{histograms1}, we can see a small reduction to the 24, 25, and 26 peaks, suggesting water has some effects on the production of HCN and hydrocarbons. The reduction in the 26/28 ratio for Experiment 6 when water is introduced is approximately a factor of 2.

Finally, we see an increase in the 41 peak in approximately half of our experimental runs. The 41 peak increase is often, but not always associated with increases in the 38 and 39 peaks. The associated peak increases can be seen in the analog spectra for Experiments 1--3 in Figure~\ref{Fig1Analog} and the histogram spectra for Experiments 2, 3, and 5 in Figure S1. The increases in the 38, 39, and 41 peaks may suggest propane (\ce{C3H8}) or acetonitrile (\ce{CH3CN}) was produced. However, given the large range of experimental relative increases for peak 38 to 41 (0.07--0.44:1) and peak 39 to 41 (0.04--5.81), we cannot use the fragmentation patterns for propane or acetonitrile to deduce which product is most likely producing these signals. We again rely on our chemical kinetic models to try to determine which of these species is the most likely source of the 41 peak increases due to the cold plasma discharge in our experiments.

\section{Results - Chemical Kinetics}

\begin{figure}[!hbtp]
\centering
\includegraphics[width=\linewidth]{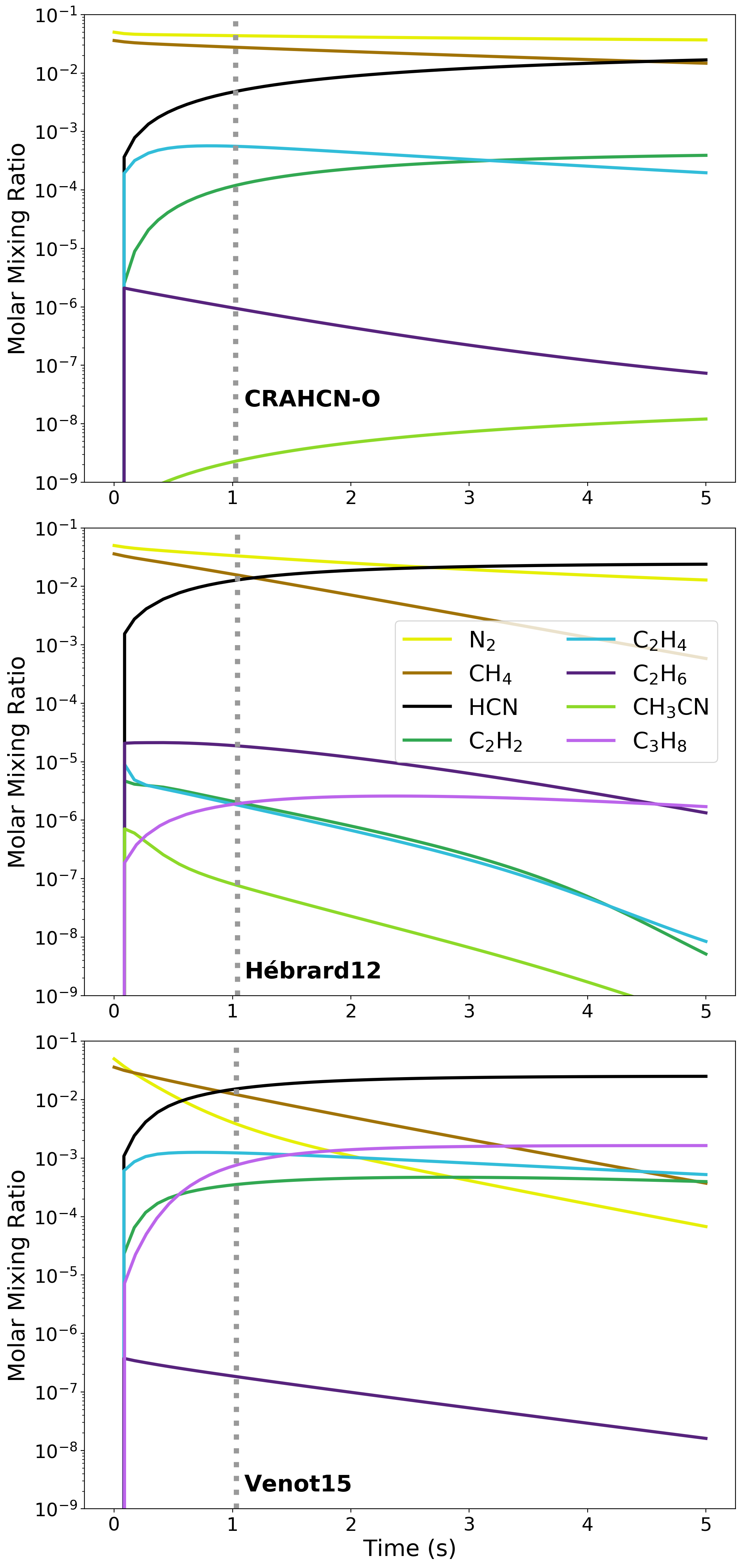}
\caption{Chemical kinetics simulations of Experiment 2 using three different chemical networks and the Druyvesteyn EEDF. The grey vertical dotted line represents the time step closest to $t$ = 1 second that is used to compare molecule concentrations across simulations. \label{Kintecus_Compare}}
\end{figure}

In Figure~\ref{Kintecus_Compare}, we display the chemical kinetics models of Experiment 2 (3.6\% \ce{CH4}) from $t$ = 0--5 seconds using each of the three chemical networks in Table~\ref{ChemicalNetworks}, and the Druyvesteyn EEDF. In Figure S2 in the Supporting Information, we display the same three models using the Maxwellian EEDF, and find that the results do not greatly differ. Perhaps the most interesting result of these models is that the concentration and temporal evolution of HCN agrees across all three networks. HCN molar mixing ratios at $t$ = 1 second differ by at most a factor of 4 across all simulations, ranging from 0.37--1.5$\times$10$^{-2}$.

The value of $t$ for our experimental setup corresponds to the length of time the gas mixture is in contact with the plasma before it flows out of the chamber and towards the RGA. Given the experimental flow rate of 5 sccm, the $\sim$2 Torr chamber pressure, and the plasma discharge volume of roughly 30 cm$^{3}$, using p$_1$v$_1$ = p$_2$v$_2$ we expect $t$ to be roughly one second. At $t$ = 1 second, we see that the second most abundant product ranges from \ce{C2H4} and \ce{C2H2} for the CRAHCN-O network simulation, \ce{C2H6} for the H{\'e}brard12 network simulation and \ce{C2H4}, \ce{C3H8}, and \ce{C2H2} for the Venot15 network simulation. The concentrations of these products range from a factor of seven to three orders of magnitude lower than the concentration of HCN.

If we compare the H{\'e}brard network to our experimental results, the dominance of the HCN concentration compared to hydrocarbon species suggests that essentially 100\% of the 26 peak would be due to HCN. However, this network solution does not accurately account for the increase in the 25 peak relative to the 26 peak, as described in the previous section. 

For the CRAHCN-O and Venot15 networks, we can roughly estimate the contribution from other species to the 26 peak by multiplying the species concentration by the species theoretical mass spectrometry fragment fraction at m/z = 26, and the species theoretical ionization cross-section at 70 eV. Then, we can divide this value for each species by the sum of this multiplier for all contributing species. The mass spectrometry fragment fractions for peak 26 are 0.17 for HCN, 1.0 for \ce{C2H2}, 0.23 for \ce{C2H6}, 0.09 for \ce{C3H8}, and 0.53 for \ce{C2H4} \citep{NISTChemWebBook_massspec}. The ionization cross-sections at 70 eV are 3.6 $\AA^{2}$ for HCN, 4.4 $\AA^{2}$ for \ce{C2H2}, 6.4 $\AA^{2}$ for \ce{C2H6}, 8.6 $\AA^{2}$ for \ce{C3H8}, and 5.1 $\AA^{2}$ for \ce{C2H4} \citep{Ionization_Cros2004,Pandya_et_al2012}. Considering these two networks, we calculate that \ce{C2H4}, \ce{C2H2}, and \ce{C3H8} could contribute up to $\sim$23--47\%, 9--15\%, and 1--4\% to peak 26, respectively.

Depending on the choice of EEDF, HCN could contribute as low as 56\% and as high as 60\% to the 26 peak increase when using the CRAHCN-O network simulations, and as low as 33\% and as high as 65\% to the 26 peak increase when using the Venot15 network simulations. To sum up these results, with the exception of the Maxwellian Venot15 simulation, our chemical kinetic models suggest that HCN is the main contributor to the 26 peak increase in our experiments due to the cold plasma discharge. We note again that low-pressure discharges are generally non-Maxwellian; therefore, we expect our Druyvesteyn EEDF results to be slightly more accurate.

The hydrocarbons that are most likely to contribute to the increase in the 24 and 25 peaks when plasma is turned on depends strongly on the chemical network. Similar to our calculations above, using CRAHCN-O network simulations, the experimental fragment patterns \citep{NISTChemWebBook_massspec} and ionization cross sections \citep{Ionization_Cros2004}, we calculate that 69--76\% of the 25 peak increase is due to \ce{C2H4} with the remainder coming from \ce{C2H2}. For the Venot15 network, the contribution from \ce{C2H4} is still the highest, at 62--70\%; however, both \ce{C2H2} and \ce{C3H8} contribute 28--37\% and 1--2\% to the signal, respectively. Finally, using the H{\'e}brard12 network simulations, we calculate that 61--62\% of the 25 peak increase is due to \ce{C2H6}, 25--26\% is due to \ce{C2H2}, 11--12\% is due to \ce{C2H4} and 1--2\% is due to \ce{C3H8}. These calculations are consistent to within 1\% when calculating the hydrocarbon contributions to the 24 peak increase for the CRAHCN-O and Venot15 network simulations, and within 14\% for the H{\'e}brard12 network simulation. Given that the Venot15 network is the most extensive network for hydrocarbon species, and there is some agreement with the CRAHCN-O network, we suggest that \ce{C2H4} and \ce{C2H2} are most likely the main contributors to the increases in the 24 and 25 peaks due to the cold plasma discharge.

We also aim to shed light on the species that produces an increase to the 41 peak in approximately half of our experiments after plasma is turned on. The two main considerations are \ce{C3H8} and \ce{CH3CN}. It is worth noting that \ce{C3H8} is not an available product in CRAHCN-O, and \ce{CH3CN} is not an available product in Venot15; therefore only the H{\'e}brard12 network can be used to compare both concentrations. Given the over an order of magnitude difference between \ce{C3H8} and \ce{CH3CN} in the H{\'e}brard12 network simulations, and the considerably high \ce{C3H8} mixing ratio for the Venot15 network simulations, we expect that the 41 peak increase due to the cold plasma discharge in our simulations is most likely due to propane rather than acetonitrile.

We find that the main difference in our models when switching from the Maxwellian and Druyvesteyn EEDF is that concentrations decrease, increase and reach steady state more rapidly.

\begin{figure}[!hbtp]
\centering
\includegraphics[width=\linewidth]{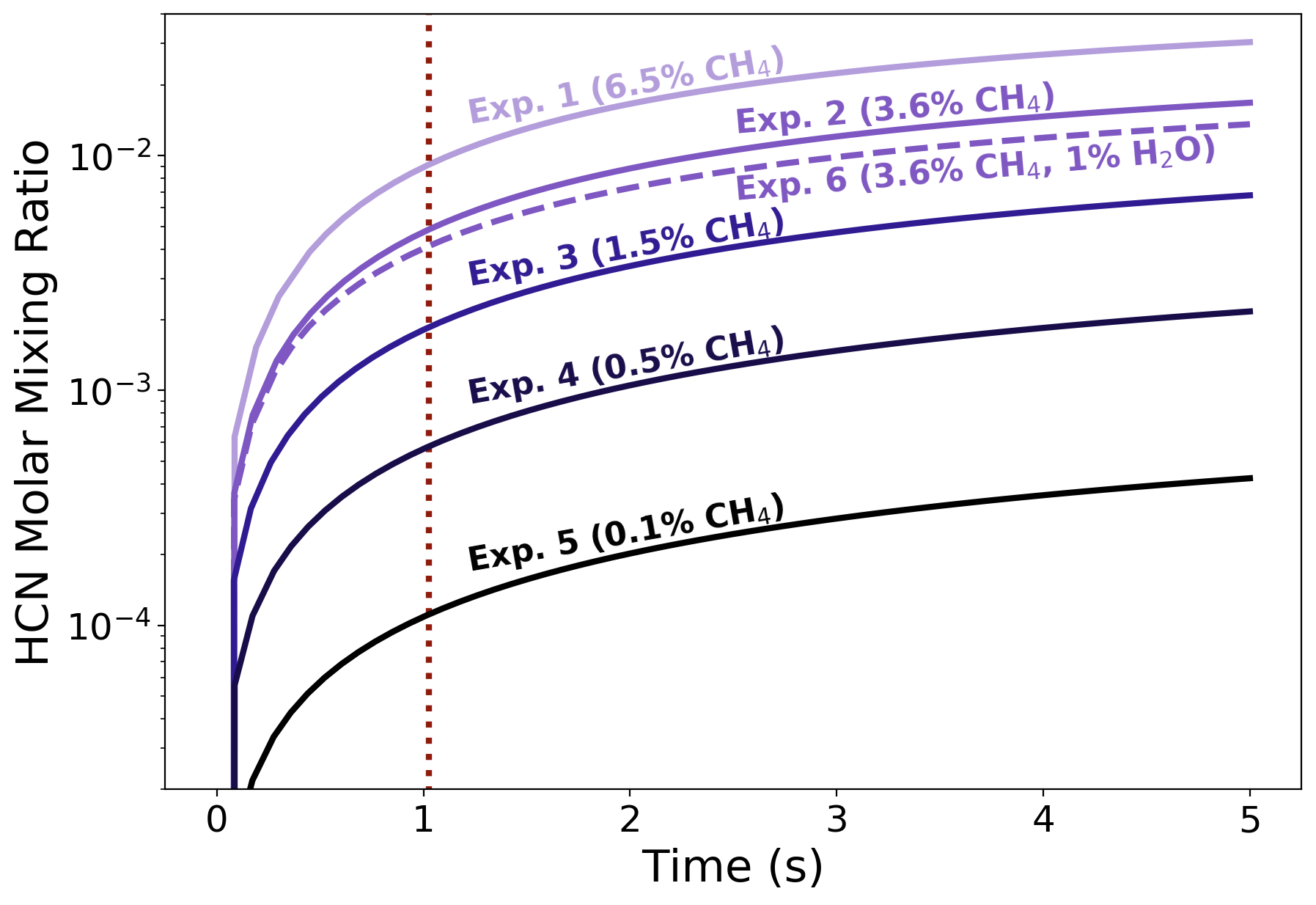}
\caption{HCN abundances from chemical kinetic simulations of Experiments 1--5 using the CRAHCN-O network and Druyvesteyn EEDF. The red vertical dotted line is roughly the time that the gas mixture is in contact with the plasma in our experiments. Note how the addition of 1\% water in the theoretical simulation of Experiment 6 decreases the HCN production by 16\% compared to Experiment 2. \label{Kintecus_HCN}}
\end{figure}

In Figure~\ref{Kintecus_HCN}, we plot the HCN abundances from our chemical kinetics models of Experiments 1--5 as well as Experiment 6 which contains water vapor. For this figure, we use the CRAHCN-O network and the Druyvesteyn EEDF. However, we note that differences in HCN abundances at $t$ = 1 second are modest across all networks and EEDFs, varying by at most a factor of 4.

The average relation between HCN and \ce{CH4} from the theoretical results of Experiments 1--5 at t = 1 second are [HCN] = 0.13$\pm$0.01 [\ce{CH4}]. These data are used for comparison with our experimental results in the following section.

\section{Results - Experiment/Theory Comparison}

\begin{figure*}[!hbtp]
\centering
\includegraphics[width=\linewidth]{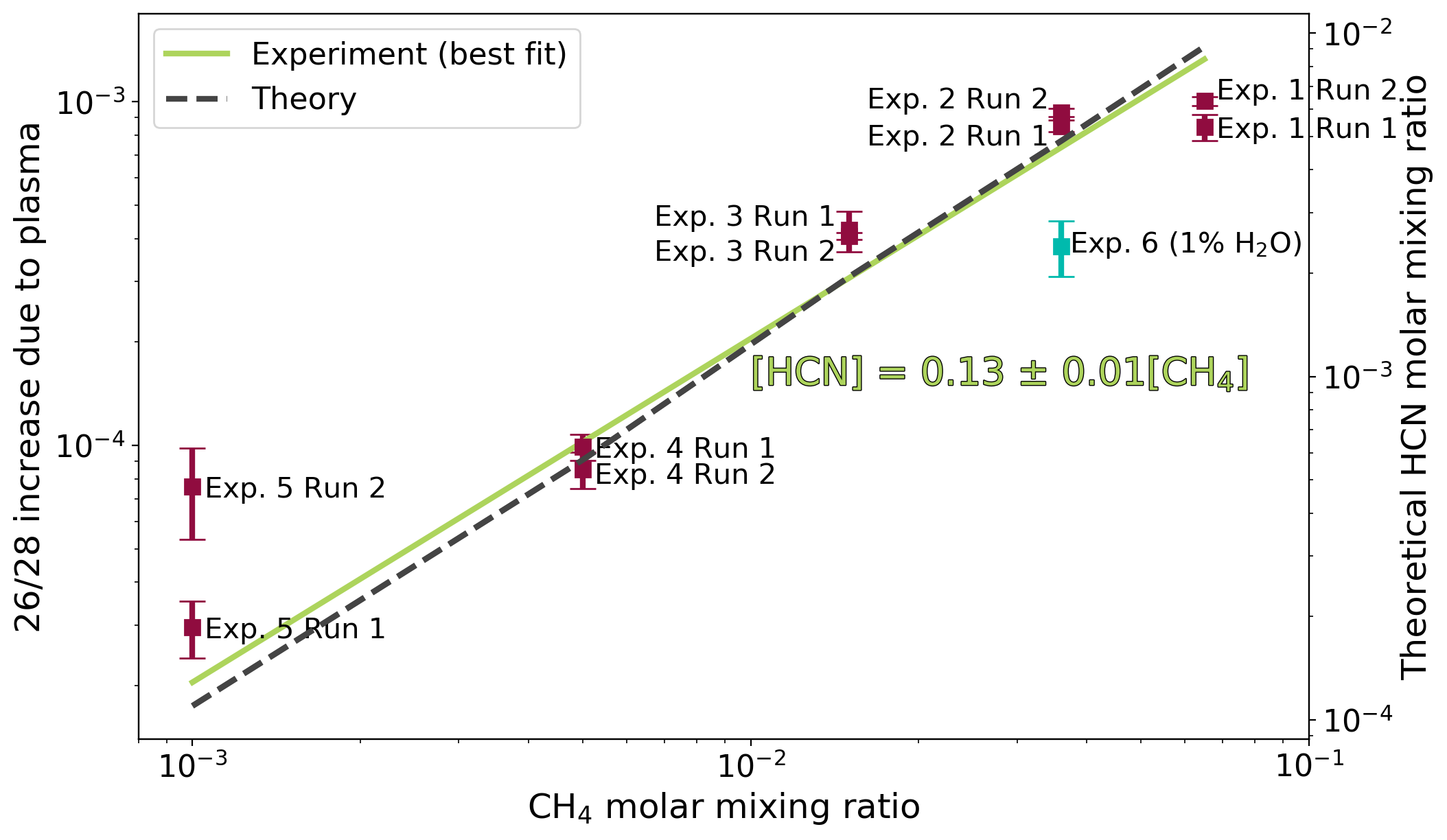}
\caption{The production of HCN as a result of turning on the cold plasma discharge for five experimental \ce{CH4} mixing ratios. HCN profuction is determined by an increase in the RGA 26 peak multiplied by a theoretical HCN contribution of 0.6 (CRACHN-O model) and normalized to the 28 peak. The results (with uncertainties) are fit to a linear function, excluding the experiment performed with water. The chemical kinetics results from Figure~\ref{Kintecus_HCN} are also displayed, calibrated to the central data point of the experimental line of best fit (\ce{CH4} = 1.5\%). The corresponding HCN molar mixing ratios are placed on the right side of the y-axis. From this we obtain a relation between HCN production and \ce{CH4} mixing ratio for our Hadean Earth experiments. \label{HCN_CH4_Exp}}
\end{figure*}

In Figure~\ref{HCN_CH4_Exp}, we display the 26/28 peak increase from cold plasma discharge, multiplied by the theoretical HCN contribution of 0.6 from our CRAHCN-O model, for our experimental Hadean atmospheres as a function of initial \ce{CH4} abundance. Then, we fit the data to a linear function to find the experimental relation between 26/28 peak increase and \ce{CH4} abundance. A linear function was chosen because, A) past spark discharge experiments suggested a fairly constant yield for HCN production for \ce{CH4} mixing ratios from 17--50\% \citep{Miller_Schlesinger1983}, which corresponds to a linear relation, and B) a linear relation fit the data reasonably well. The calculated relation is [(26/28)$_{plasma}$ - (26/28)$_{gas}$] $\times$ 0.6 = 0.02[\ce{CH4}].


Next, we place the results of our coupled theoretical models on the same plot, by calibrating to the central point on the linear relation. In other words, we assume that the 26/28 peak increase at [\ce{CH4}] = 0.015 is due to the theoretical HCN concentration produced in our coupled chemical kinetics model in Figure~\ref{Kintecus_HCN}. Each point on the theoretical 26/28 peak increase curve is calculated from the HCN concentration from the corresponding chemical kinetics model. The relation between HCN production and \ce{CH4} mixing ratio from these theoretical models---calculated in Section 4---is [HCN] = 0.13 $\pm$ 0.01[\ce{CH4}].


Next, on the right side of the y-axis, we label the corresponding HCN concentrations for each point on the theoretical 26/28 peak increase curve. In doing this, we have created a relation between our experimental 26/28 peak increases and theoretical HCN production, linked by a conversion factor of 5.37.

The validity of using the theoretical relation for the experimental results can be analyzed by looking at the agreement between the shapes of the two curves. We find good agreement between the theoretical HCN curve and the experimental 26/28 peak increase curve, with a maximum difference of 15\% at [\ce{CH4}] = 0.001. Given all possible calibration points from [\ce{CH4}] = 0.001--0.065, the maximum difference can increase up to 22\%. These results suggest that roughly 13 $\pm$ 1\% of \ce{CH4} is converted into HCN in our experiments.

We emphasize that the relation between HCN production and \ce{CH4} mixing ratio derived from our experiments and theory is based on plasma modeling results using the CRAHCN-O network and Druyvesteyn EEDF. We think this network is the most accurate of the three networks for modeling HCN production, as it contains nearly 100 reactions previously undiscovered in the literature that were calculated using consistent quantum chemistry and statistical mechanics methods that have gone through rigorous validations in comparison to experimental data \citep{Reference598,Pearce2020a,Pearce2020b,Pearce_et_al2022a}. This network has also been validated with observational data when modeling HCN production in the atmosphere of Titan \citep{Pearce2020a}. When using the H{\'e}brard12 and Venot networks and the Druyvesteyn EEDF, the HCN relations increase to [HCN] = 0.43 $\pm$ 0.06 and 0.41 $\pm$ 0.02[\ce{CH4}], respectively. These yields are quite high compared to past experiments (0.09–0.0135 \citep{Miller_Schlesinger1983}), which may suggest these networks are less accurate for modeling HCN production and destruction. In the sensitivity analysis in Section 6 we analyze these networks in detail and suggest potential reasons for the inaccuracies.

The largest discrepancy between the two runs of each experiment are for Experiment 5, where there is approximately a factor of 2.5 difference between Runs 1 and 2. For Experiment 5, we made the \ce{CH4} mixing ratio as low as possible while still remaining within the detection limit for changes to the 26 peak. Given the m/z = 26 signal was just above the background for these runs, we expect the uncertainty of the Experiment 5 results to be the highest of all experiments.

Lastly, we plot the experiment with 1\% water and find that the addition of water leads to a 50\% reduction in HCN production. This experimental reduction in HCN due to water is about 34\% larger than the theoretical reduction in our chemical kinetic models.

\section{Results - Sensitivity Analyses}

In order to understand which reactions are critical for the production and destruction of HCN in our Hadean Earth experiments, we run sensitivity analyses for Experiment 2 and Experiment 6 (which only differ by the absence or presence of 1\% water), for each of the three chemical networks employed in this study. The sensitivity analyses involve removing one reaction at a time, running the chemical kinetic simulation, and comparing the difference to the HCN abundance at $t$ = 1 second with the reference simulation. If the HCN abundance differs by less than the threshold (chosen to be 3\%), than the reaction is discarded. At the end of the simulation, a list of $\sim$15--25 dominant reactions for HCN production and destruction remain.

In Table~\ref{Sensitivity_Analyses_Table}, we list the dominant HCN source and sink reactions for Experiment 2, for each of the three networks. We also include the percent difference to the HCN abundance (i.e. the relative importance of the reaction) due to the removal of the reaction from the network.

\begin{table*}[ht!]
\centering
\caption{The results of the HCN sensitivity analyses for Experiment 2 for the three reaction networks employed in this study, using the Druyvesteyn EEDF for plasma chemistry. The top 10 dominant sources and sinks are included for percent increases/decreases $>$ 3\%. In bold are reactions dominant in all three network simulations. In purple are reactions dominant in two of the three network simulations. In orange are reactions that are a dominant source in one network simulation and a dominant sink in another network simulation. \label{Sensitivity_Analyses_Table}} 
\begin{tabular}{llclc}
\\
\multicolumn{1}{c}{Network} &  
\multicolumn{1}{l}{Source Reaction} & 
\multicolumn{1}{l}{Percent Dec.} &
\multicolumn{1}{l}{Sink Reaction} &
\multicolumn{1}{l}{Percent Inc.} 
\\[+2mm] \hline \\[-2mm]
CRAHCN-O & {\bf \ce{N2 + e^- -> ^4N + ^4N + e^-} }& 100\% & \textcolor{purps}{\ce{CH3 + H + M -> CH4 + M}} & 303\% \\
  & \textcolor{purps}{\ce{CH4 + e^- -> ^3CH2 + H2 + e^-}} & 59\% & {\bf \ce{H2 + e^- -> H + H + e^-}} & 202\% \\
    & \textcolor{purps}{\ce{C + H2 + M -> ^3CH2 + M}} & 51\% & \ce{^4N + NH -> N2 + H} & 119\% \\
 & \textcolor{purps}{\ce{^4N + CH3 -> H2CN + H}} & 49\% & \ce{^4N + H + M -> NH + M} & 98\% \\
  & \ce{NH + H -> ^4N + H2} & 46\% & \ce{CH + H2 + M -> CH3 + M} & 34\% \\
 &  {\bf \ce{CH4 + e^- -> CH3 + H + e^-}} & 35\% &  {\bf \ce{^3CH2 + H -> CH + H2} }& 24\% \\
  & \textcolor{purps}{\ce{H + H + M -> H2 + M}} & 33\% & \ce{^3CH2 + H + M -> CH3 + M} & 5\% \\
 &  \ce{C2H4 + ^4N -> HCN + CH3} & 30\% &  & \\
 & \textcolor{orang}{\ce{CH4 + CH -> C2H4 + H}} & 16\% &  &\\
 & \textcolor{purps}{\ce{^4N + ^3CH2 -> HCN + H}} & 11\% &  & \\
  & & & & \\
 H{\'e}brard12 & {\bf \ce{N2 + e^- -> ^4N + ^4N + e^-} } & 100\% & {\bf \ce{^3CH2 + H -> CH + H2} } & 35\% \ \\
  & {\bf \ce{CH4 + e^- -> CH3 + H + e^-}} & 63\% & \textcolor{purps}{\ce{CH3 + H + M -> CH4 + M}} & 27\% \\
  & \textcolor{purps}{\ce{^4N + CH3 -> H2CN + H}} & 45\% & \textcolor{purps}{\ce{CH + H -> C + H2}} & 22\% \\
   & \textcolor{purps}{\ce{H + H + M -> H2 + M}} & 23\% & \ce{HCN + C2 -> C3N + H} & 21\% \\
     & \ce{C2H3 + H + M -> C2H4 + M} & 16\% & \ce{H + C2H3 -> C2H2 + H2} & 21\% \\
     & \ce{C2N + H -> HCN + C} & 13\% & \ce{CN + ^4N -> N2 + C} & 20\% \\
     & \textcolor{purps}{\ce{C + H2 + M -> ^3CH2 + M}} & 10\% & \ce{C + C2H4 -> C3H3 + H}  & 14\% \\
     & \ce{H + C3H5 + M - C3H6 + M} & 67\% & \ce{^4N + C3 -> CN + C2} & 14\% \\
     & \ce{H2CN + H -> HCN + H2} & 7\% & {\bf \ce{H2 + e^- -> H + H + e^-}} & 14\% \\
 &  \ce{H + CH3C2H + M -> C3H5 + M} & 6\% & \ce{C + C2H2 -> C3 + H2} & 10\% \\
  & & & & \\
 Venot15 & {\bf \ce{N2 + e^- -> ^4N + ^4N + e^-} } & 100\% &  {\bf \ce{^3CH2 + H -> CH + H2} } & 25\% \\
 & {\bf \ce{CH4 + e^- -> CH3 + H + e^-}} & 58\% & {\bf \ce{H2 + e^- -> H + H + e^-}} & 18\% \\
 & \textcolor{purps}{\ce{CH4 + e^- -> ^3CH2 + H2 + e^-}}  & 22\% & \textcolor{orang}{\ce{CH4 + CH -> C2H4 + H}} & 18\% \\
 & \ce{^4N + CH -> CN + H} & 25\% & \ce{H2CN + ^4N -> N2 + ^3CH2} & 13\% \\
 & \ce{C3H2 + H -> C2H3} & 20\% & \ce{CN + H2 -> HCN + H} & 8\% \\
 & \ce{C3H3 + H -> C3H2 + H2} & 20\% & \textcolor{purps}{\ce{CH + H -> C + H2}} & 8\% \\
 & \ce{CH + C2H2 -> C3H2 + H} & 20\% &  & \\
 & \textcolor{purps}{\ce{^4N + ^3CH2 -> HCN + H}} & 14\% & & \\
 & \ce{CN + CH4 -> HCN + CH3} & 6\% & & \\
 & \ce{CH3 + C -> C2H2 + H} & 3\% & & \\
\\[-2mm] \hline
\end{tabular}
\end{table*}

We find that the number one source is consistent across all three network simulations. \ce{N2 + e^- -> ^4N + ^4N + e^-} is the most critical source reaction, as there is no other efficient route to break the triple bond of \ce{N2} to produce nitrogen radicals that react to produce HCN. Furthermore, a methane dissociation reaction, either \ce{CH4 + e^- -> ^3CH2 + H2 + e^-} or \ce{CH4 + e^- -> CH3 + H + e^-} is consistent as the second most important source reaction across all networks. This is also not surprising, as HCN is known to be produced via reactions involving nitrogen and methane radicals.

The key direct source of HCN varies across network simulations. For CRAHCN-O, the number one direct route to HCN is \ce{^4N + CH3 -> H2CN + H}---which proceeds most rapidly to HCN via \ce{H2CN + H -> HCN + H2}. The second step of this reaction does not appear in the Table~\ref{Sensitivity_Analyses_Table} because although it is the dominant subsequent reaction, there are other efficient replacement reactions when it is removed, e.g., \ce{^4N + H2CN -> HCN + NH}. Both \ce{C2H4 + ^4N -> HCN + CH3} and \ce{^4N + ^3CH2 -> HCN + H} are also important contributors to HCN production for this network. For H{\'e}brard12, the most efficient direct route to HCN is also \ce{^4N + CH3 -> H2CN + H -> HCN + H2}. However, \ce{C2N + H -> HCN + C} contributes a significant amount to HCN production as well. In the Venot15 network simulation, the dominant direct route to HCN is \ce{^4N + CH -> CN + H}---which proceeds to HCN via \ce{CN + CH4 -> HCN + CH3}. The second dominant source of HCN for this network is \ce{^4N + ^3CH2 -> HCN + H}. We note that the rate coefficient for the dominant HCN production reaction in the Venot15 simulation is approximately a factor of two higher than the value suggested by the average of experiments \citep{Reference572,Reference570,Reference571}.

Given the lack of experimental studies for the \ce{C2N + H -> HCN + C} pathway for HCN production in the H{\'e}brard12 network simulation, we analyze this reaction in detail using quantum chemistry. \citet{2012AA...541A..21H} were the first to estimate a rate coefficient for this reaction for use in the chemical modeling of Titan's atmosphere. They used capture rate theory to estimate a barrierless upper limit rate coefficient of 2.0$\times$10$^{-10}$ cm$^3$ s$^{-1}$. \citet{Takahashi_Takayanagi2006} studied this reaction with quantum chemistry, and found the pathway to involve four transition states, i.e., \ce{C2N + H -> ^3HCCN -> ^3HCCN_{isomer} -> ^3HCNC -> HCN + C}. Their calculations show that transition states 2, 3, and 4 are found to have barriers of 58.2, 4.9, and 66.7 kcal/mol at the B3LYP/6-311++G(d,p) level of theory, respectively. We go one step further to analyze this reaction, by 1) remodeling the transition state chemistry for this reaction using the Gaussian 09 software package \citep{g09}, and 2) calculating the reaction rate coefficient with canonical variational transition state theory \citep{Reference534,Reference598,Pearce2020a,Pearce2020b} using the Becke--Half-and-Half--Lee--Yang--Parr (BHandHLYP) density functional and the augmented correlation-consistent polarized valence double-$\zeta$ (aug-cc-pVDZ) basis set \citep{Reference594,Reference595,Dunning1989,KendallDunning1992,WoonDunning1993}. Given this reaction has multiple steps, we use mechanistic modeling to calculate the steady state solutions to the kinetic rate equations in order to obtain the overall reaction rate coefficient (see supplement of \citet{Reference598,Pearce2020b} for more detail).

We originally developed this accurate and feasible method to calculate rate coefficients for reactions relevant to HCN chemistry in the atmospheres of early Earth and Titan. We have used this method in the past to calculate hundreds of rate coefficients, including 93 reactions that were previously undiscovered prior to our exploratory studies \citep{Reference598,Pearce2020a,Pearce2020b,Pearce_et_al2022a}. Our validation studies with experiments show that reaction rate coefficients calculated using this method most frequently land within a factor of two of experimental values, and are always within about an order of magnitude of experimental values \citep{Pearce2020a,Pearce2020b}. Contrary to the barrierless estimate by \citet{2012AA...541A..21H}, we find that this reaction proceeds through multiple barriers, consistent with the quantum chemistry results by \citet{Takahashi_Takayanagi2006}. Our calculated rate coefficient is 2.6$\times$10$^{-50}$ cm$^3$ s$^{-1}$ - which can be considered negligible. Furthermore, if we instead use the energy barriers calculated by \citet{Takahashi_Takayanagi2006} for our calculations, which use a density functional known to underestimate barrier heights \citep{Reference598,Pearce2020a,Pearce2020b}, the calculated rate coefficient is still negligible, at 8.2$\times$10$^{-45}$ cm$^3$ s$^{-1}$. Given these results, we conclude that \ce{C2N + H -> HCN + C} is not a dominant source of HCN in our Hadean Earth experiments.

Considering the CRAHCN-O and Venot15 networks, there is a discrepancy in the number one direct pathway to HCN. The main reason for this is that the Venot15 network has an extensive set of reactions that build larger hydrocarbons using smaller hydrocarbons (e.g., \ce{C2H4}) and methane fragments (e.g., \ce{CH3} and \ce{^3CH2}). The competition between this extensive set of reactions and HCN reactions for hydrocarbons and methane fragments shifts the dominance of direct HCN production to cyanide routes. One caveat to using the extensive Venot15 network is that, due to lack of data, we do not include electron impact dissociation reactions for the hundreds of large hydrocarbons in these chemical kinetics simulations. It may be the case that including such data would recycle more \ce{CH3} and \ce{^3CH2} fragments back into the gas mixture, and the methane fragment routes for HCN production would become more dominant as with the CRAHCN-O network simulations. This, however, remains uncertain.

Overall, we see that the four main direct routes to HCN in the dry simulations are \ce{^4N + CH3 -> H2CN + H -> HCN + H2}, \ce{C2H4 + ^4N -> HCN + CH3}, \ce{^4N + CH -> CN + H} followed by \ce{CN + CH4 -> HCN + CH3}, and \ce{^4N + ^3CH2 -> HCN + H}. Ironically, \ce{CN + H2 -> HCN + H} is actually a slight sink reaction in the Venot15 simulation. The reason is because this reaction produces HCN at the expense of producing H atoms, which destroy HCN precursor \ce{^3CH2}. The dominant cyanide route to HCN, i.e., \ce{CN + CH4 -> HCN + CH3}, produces an HCN precursor (\ce{CH3}) rather than H atoms.

The most critical sink reactions are \ce{CH3 + H + M -> CH4 + M} and \ce{^3CH2 + H -> CH + H2}. The first sink removes \ce{CH3}, which is a key reactant for \ce{^4N + CH3 -> H2CN + H}. The second sink removes \ce{^3CH2}, which is a reactant for \ce{^4N + ^3CH2 -> HCN + H}. Another key sink across all network simulations is \ce{H2 + e^- -> H + H + e^-}. One reason for this is that the source reaction \ce{C + H2 + M -> ^3CH2 + M} requires \ce{H2} to make a key precursor to HCN. Another reason is that H atoms efficiently destroy \ce{^3CH2} via the sink reaction \ce{^3CH2 + H -> CH + H2}. Other key sinks in the list are involved in the destruction of precursors to HCN, such as \ce{CN}, \ce{^4N}, and \ce{H2CN}. Interestingly, \ce{CH4 + CH -> C2H4 + H} is a key sink reaction in the Venot15 simulation, and a key source reaction in the CRAHCN-O simulation. This is due to the differences in the key direct sources of HCN between these simulations. \ce{C2H4} is a key direct source of HCN in the CRAHCN-O simulation via \ce{C2H4 + ^4N -> HCN + CH3}, whereas \ce{CH4} is a direct source of HCN in the Venot15 simulation via \ce{CN + CH4 -> HCN + CH3}.


In Table~\ref{Sensitivity_Analyses_Table_Water}, we display the dominant source and sink reactions for HCN from a sensitivity analysis of Experiment 6 - which contains a 1\% molar mixing ratio of water. Labelled in blue are all the reactions that were not previously dominant in the dry sensitivity analyses of Experiment 2 in Table~\ref{Sensitivity_Analyses_Table}.

\begin{table*}[ht!]
\centering
\caption{The results of the HCN sensitivity analyses for Experiment 6 for the three reaction networks employed in this study, using the Druyvesteyn EEDF for plasma chemistry. The top 10 dominant sources and sinks are included for percent increases/decreases $>$ 3\%. In bold are reactions dominant in all three network simulations. In blue are reactions that are not dominant in the corresponding simulation without water in Table~\ref{Sensitivity_Analyses_Table}. \label{Sensitivity_Analyses_Table_Water}} 
\begin{tabular}{llclc}
\\
\multicolumn{1}{c}{Network} &  
\multicolumn{1}{l}{Source Reaction} & 
\multicolumn{1}{l}{Percent Dec.} &
\multicolumn{1}{l}{Sink Reaction} &
\multicolumn{1}{l}{Percent Inc.} 
\\[+2mm] \hline \\[-2mm]
CRAHCN-O & {\bf \ce{N2 + e^- -> ^4N + ^4N + e^-} }& 100\% & \ce{CH3 + H + M -> CH4 + M} & 354\% \\
  & \ce{CH4 + e^- -> ^3CH2 + H2 + e^-} & 58\% & {\bf \ce{H2 + e^- -> H + H + e^-}} & 227\% \\
    & {\bf \ce{^4N + CH3 -> H2CN + H}} & 56\% & \ce{^4N + H + M -> NH + M} & 105\% \\
 & \ce{C + H2 + M -> ^3CH2 + M} & 45\% & \ce{^4N + NH -> N2 + H} & 98\% \\
  & \ce{NH + H -> ^4N + H2} & 45\% & \ce{CH + H2 + M -> CH3 + M} & 41\% \\
 &  {\bf \ce{CH4 + e^- -> CH3 + H + e^-}} & 40\% & {\bf \ce{^3CH2 + H -> CH + H2} }& 28\% \\
  & {\bf \ce{H + H + M -> H2 + M}} & 36\% & \textcolor{blue}{\ce{H2O + e^- -> OH + H + e^-}} & 17\% \\
 & \ce{C2H4 + ^4N -> HCN + CH3} & 16\% & \textcolor{blue}{\ce{OH + ^4N -> NO + H}} & 17\% \\
 & \ce{^4N + ^3CH2 -> HCN + H} & 14\% & \ce{^3CH2 + H + M -> CH3 + M} & 6\% \\
 & \ce{CH4 + CH -> C2H4 + H} & 13\% & &  \\
  & & & & \\
  H{\'e}brard12 & {\bf \ce{N2 + e^- -> ^4N + ^4N + e^-} } & 100\% & {\bf \ce{^3CH2 + H -> CH + H2} } & 35\% \ \\
  & {\bf \ce{CH4 + e^- -> CH3 + H + e^-}} & 64\% &\ce{CH3 + H + M -> CH4 + M} & 26\% \\
  & {\bf \ce{^4N + CH3 -> H2CN + H}} & 45\% & \ce{CH + H -> C + H2} & 20\% \\
   & {\bf \ce{H + H + M -> H2 + M}} & 22\% & \ce{HCN + C2 -> C3N + H} & 20\% \\
     & \ce{C2H3 + H + M -> C2H4 + M} & 14\% & \ce{CN + ^4N -> N2 + C} & 18\% \\
     &\ce{C2N + H -> HCN + C}  & 13\% & \ce{H + C2H3 -> C2H2 + H2} & 18\% \\
     & \ce{C + H2 + M -> ^3CH2 + M} & 9\% & {\bf \ce{H2 + e^- -> H + H + e^-}} & 15\% \\
     & \ce{H2CN + H -> HCN + H2} & 7\% & \ce{C + C2H4 -> C3H3 + H} & 13\% \\
     & \ce{H + C3H5 + M - C3H6 + M} & 7\% & \ce{^4N + C3 -> CN + C2} & 12\% \\
 &  \ce{H + CH3C2H + M -> C3H5 + M} & 6\% & \ce{C + C2H2 -> C3 + H2}  & 9\% \\
  & & & & \\
 Venot15 & {\bf \ce{N2 + e^- -> ^4N + ^4N + e^-} } & 100\% &  {\bf \ce{^3CH2 + H -> CH + H2} } & 38\% \\
 & {\bf \ce{CH4 + e^- -> CH3 + H + e^-}} & 47\% & {\bf \ce{H2 + e^- -> H + H + e^-}} & 31\% \\
 & \ce{^4N + CH -> CN + H} & 37\% & \ce{H2CN + ^4N -> N2 + ^3CH2} & 29\% \\
 & {\bf \textcolor{blue}{\ce{H + H + M -> H2 + M}}} & 23\% & \textcolor{blue}{\ce{C + NO -> CO + ^4N}} & 20\% \\
 & \ce{CH4 + e^- -> ^3CH2 + H2 + e^-} & 15\% & \ce{CH + H -> C + H2} & 18\% \\
 & \ce{^4N + ^3CH2 -> HCN + H}& 13\% & \textcolor{blue}{\ce{C3H3 + ^3O -> C2H + H2CO}} & 15\% \\
 & \textcolor{blue}{\ce{H2CN + H -> HCN + H2}} & 8\% & \ce{CN + H2 -> HCN + H} & 10\% \\
 & \ce{CN + CH4 -> HCN + CH3} & 7\% & \ce{CH4 + CH -> C2H4 + H} & 10\% \\
 & {\bf \textcolor{blue}{\ce{^4N + CH3 -> HCN + H2}}} & 3\% & \textcolor{blue}{\ce{H2O + e^- -> OH + H + e^-}} & 7\% \\
 & & & \textcolor{blue}{\ce{^4N + OH -> NO + H}} & 4\% \\
\\[-2mm] \hline
\end{tabular}
\end{table*}

Only in the CRAHCN-O and Venot15 network simulations are there differences to the dominant source and/or sink reactions with the addition of water.

In the CRAHCN-O network, the three direct sources of HCN remain the same when water is added. For sinks, two main new reactions create an approximate 16\% reduction in HCN by removing \ce{^4N} atoms that would otherwise be used to produce HCN. These reactions are \ce{H2O + e^- -> OH + H + e^-} followed by \ce{OH + ^4N -> NO + H}.

There are three new dominant source reactions and four new dominant sink reactions for the Venot15 network simulation when water is added. The most dominant new source reaction is \ce{H + H + M -> H2 + M}, which removes H atoms that break down \ce{^3CH2} in the number one sink reaction. The most dominant new sink reaction is \ce{C + NO -> CO + ^4N}, which oxidizes carbon atoms, making the carbon energetically unavailable for the production of HCN. The dominant direct source of HCN, i.e., \ce{^4N + CH -> CN + H}, followed by \ce{CN + CH4 -> HCN + CH3}, remains the same when water is added. The new reaction \ce{^4N + CH3 -> HCN + H2} joins the list of dominant direct sources of HCN. This reaction is the same as the number one dominant source reaction in the CRAHCN-O and H{\'e}brard12 networks, i.e., \ce{^4N + CH3 -> H2CN + H -> HCN + H2}. The quantum chemistry shows that this reaction always proceeds through \ce{H2CN} \citep{Reference598}.

In summary, when water is introduced to the simulation, the four main direct routes to HCN remain dominant, i.e., \ce{^4N + CH3 -> H2CN + H -> HCN + H2}, \ce{^4N + CH -> CN + H} followed by \ce{CN + CH4 -> HCN + CH3}, \ce{C2H4 + ^4N -> HCN + CH3}, and \ce{^4N + ^3CH2 -> HCN + H}.

\section{Discussion and Conclusions}

Several early Earth models have revealed the connection between atmospheric HCN production and \ce{CH4} abundance \citep{Pearce_et_al2022a,Zahnle_et_al2020,2019Icar..329..124R,2011EPSL.308..417T,Reference591}. Moreover, experiments simulating lightning-based chemistry in the Hadean Earth atmosphere have found nearly constant 9--13.5\% yields for very high (17--50\%) atmospheric \ce{CH4} mixing ratios \citep{Miller_Schlesinger1983}. We are the first to use simulate upper atmospheric UV chemistry for a range of plausible Hadean \ce{CH4} abundances, to find a relation between atmospheric \ce{CH4} content and HCN production at the earliest habitable stages of our planet.

For our experimental setup, we see a relation between HCN production and \ce{CH4} molar mixing ratio that follows the function [HCN] = 0.13 $\pm$ 0.01[\ce{CH4}]. Our relation, which extends from [\ce{CH4}] = 0.001–0.065, agrees well with spark discharge experiments performed by \citet{Miller_Schlesinger1983} for [\ce{CH4}] = 0.17--0.5, which they find to be [HCN] = 0.09--0.135[\ce{CH4}]. How do these experimental relations relate to HCN production from shortwave UV chemistry on the Hadean Earth?


In our previous study \citet{Pearce_et_al2022a}, we performed 1D Hadean Earth atmospheric photochemical simulations for \ce{H2}-dominant atmospheres, with methane abundances in the ppm-range. We saw peak HCN production from UV chemistry occurred at altitudes of 500 km and pressures of $\sim$0.0015 Torr. Peak HCN production in these 1D atmospheric models follows the relation [HCN] = 0.73--0.91[\ce{CH4}] after 30--50 million years. The main difference between the 1D atmospheric models and our experiments is gas exposure time to UV/plasma. We see in our 0D chemical kinetics models in Figure~\ref{Kintecus_HCN} that HCN has not reached steady state by 1 second. Thus, we expect that longer exposure times such as what would occur in the Hadean atmosphere, would result in higher HCN yields.

HCN produced in the upper Hadean atmosphere is turbulently mixed downwards to the troposphere where it can rain out into favorable environments for the emergence of life such as warm little ponds \citep{Pearce_et_al2022a}. Early Earth photochemical models suggest the HCN abundances near the surface are approximately 2--3 orders of magnitude lower than the peak abundances produced in the upper atmosphere \citep{Pearce_et_al2022a,2011EPSL.308..417T,Reference591}.

We find that water has a slight impact on HCN production. Experimentally, we see a reduction of about 50\% in HCN after adding a molar mixing ratio of $\sim$1\% water in Experiment 6. Using the CRAHCN-O network, we see a theoretical 16\% reduction in HCN due to the addition of water. The discrepancy is possibly due to missing reactions in our network, such as ion-based reactions. Although we worked hard to discover all the previously unknown neutral reactions related to \ce{H2O}, its dissociation products, and HCN in \citet{Pearce2020b}, we still do not include ion chemistry, and likely several unstudied electron impact dissociation reactions. In any case, a 34\% discrepancy is not very large and hints that the CRAHCN-O network does a reasonable job at simulating HCN chemistry in the presence of \ce{H2O}. We also note that the upper atmosphere is unlikely to have water vapor mixing ratios as high as 1\%; therefore the reduction of HCN production due to the presence of water in the upper Hadean atmosphere is likely much less than 50\%.

We find that the key direct sources of HCN are the reactions (A) \ce{^4N + CH3 -> H2CN + H -> HCN + H2}, (B) \ce{^4N + CH -> CN + H} followed by \ce{CN + CH4 -> HCN + CH3}, (C) \ce{C2H4 + ^4N -> HCN + CH3}, and (D) \ce{^4N + ^3CH2 -> HCN + H}. The first two reactions are consistent with two of the four dominant sources of HCN in Titan's atmosphere \citep{Pearce2020a}. The dominant sink \ce{^3CH2 + H -> CH + H2} is also one of the dominant sinks in models of Titan's atmosphere \citep{Pearce2020a}. We find that the addition of water to our simulations does not have a great impact on the dominant sources and sinks for HCN. The main source of the 16\% theoretical reduction in HCN is due to the two reactions \ce{H2O + e^- -> OH + H + e^-} followed by the \ce{^4N} removal reaction, \ce{OH + ^4N -> NO + H}.

Since HCN production scales roughly linearly with \ce{CH4} abundance in the upper Hadean atmosphere, a rich \ce{CH4} (e.g., $>$5\%) reducing atmosphere is the most favorable for prebiotic chemistry. Paradoxically, an atmosphere too rich in \ce{CH4} and \ce{H2} may also be uninhabitable, given the greenhouse effects maintained by \ce{H2-H2} collisional induced absorption and \ce{CH4} absorption in the infrared \citep{Pearce_et_al2022a}. The atmosphere requires the depletion of \ce{CH4} to cool, which reduces HCN production. If HCN is best produced in the upper atmosphere when the surface is too hot for liquid water, then how might it persist until the surface cools to habitable temperatures?

One solution might be for HCN to be incorporated into solids during this hot \ce{CH4}-rich phase. For example, organic hazes produced from discharge experiments of gas mixtures with 2--5\% \ce{CH4} have been shown to contain several biomolecules including all five canonical nucleobases and several proteinogenic amino acids \citep{2012AsBio..12..809H,Sebree_et_al2018}. Perhaps HCN and biomolecules could have survived in organic hazes while the Hadean atmosphere was hot, and later been incorporated into surface ponds after sufficient cooling. Organic hazes also have a cooling effect on the atmosphere, and may increase the cooling rate of the surface \citep{Reference124}. In future experiments, we aim to explore the production and organic composition of organic hazes and exposing them to the hot conditions present in a \ce{CH4}-rich Hadean troposphere. Such solids may have been a key source of biomolecule feedstock to warm little ponds on early Earth.

\section*{Acknowledgements}

We thank the anonymous referees, whose comments improved this work. B.K.D.P. is supported by the NSERC Banting Postdoctoral Fellowship. C.H. acknowledges NASA Grant 80NSSC20K0271.


\beginappendixA
\section*{Supporting Information}

In Table~\ref{addedtwobodychemistry}, we display the 53 chemical reactions added to the CRAHCN-O chemical network for our 0D chemical kinetics models of the Hadean Earth. CRAHCN-O was originally developed to accurately model HCN and \ce{H2CO} in atmospheres dominated by any of \ce{H2}, \ce{CH4}, \ce{H2O}, \ce{CO2}, and \ce{N2}. We added these new reactions in order to better estimate the production of hydrocarbons and a few other potential molecules of interest including \ce{NH3} and \ce{NO}.

\setlength\LTcapwidth{\textwidth}
\begin{longtable*}{lcccc}
\caption{New two-body reactions added to CRAHCN-O for our Hadean Earth atmospheric models, and their experimental or theoretical Arrhenius coefficients. These reactions are added to better estimate the production of hydrocarbons and other potential species of interest such as \ce{NO} and \ce{NH3}. For complete network, see tables in \citet{Pearce2020b,Pearce2020a,Pearce_et_al2022a}. The Arrhenius expression is $k(T) = \alpha \left(\frac{T}{300}\right)^{\beta} e^{-\gamma/T}$. \label{addedtwobodychemistry}} \\
Reaction Equation & $\alpha$ & $\beta$ & $\gamma$ & Source(s) \\ \hline \\[-2mm]
\ce{C2H + H2 -> C2H2 + H} & 2.5$\times$10$^{-11}$ & 0 & 1560 & \citet{Reference451}\\    
\ce{C2H + OH -> C2H2 + ^3O} & 3.0$\times$10$^{-11}$ & 0 & 0 & \citet{Reference509}\\ 
\ce{C2H + OH -> CO + ^3CH2} & 3.0$\times$10$^{-11}$ & 0 & 0 & \citet{Reference509}\\
\ce{C2H + HO2 -> OH + HCCO} & 3.0$\times$10$^{-11}$ & 0 & 0 & \citet{Reference509}\\ 
\ce{C2H + HCN -> HC3N + H} & 5.3$\times$10$^{-12}$ & 0 & 769 & \citet{Hoobler_Leone1997},\\
 & & & & \citet{Hebrard_et_al2009} \\
\ce{C2H2 + ^3O -> CO + ^3CH2} & 3.5$\times$10$^{-12}$ & 1.5 & 850 & \citet{Cvetanovic1987}\\ 
\ce{C2H2 + CN -> HC3N + H} & 2.3$\times$10$^{-10}$ & 0 & 0 & \citet{Gannon_et_al2007}\\ 
\ce{C2H2 + CN -> HCN + C2H} & 2.2$\times$10$^{-10}$ & 0 & 0 & \citet{Sayah_et_al1988}\\ 
\ce{C2H4 + ^3O -> HCO + CH3} & 8.9$\times$10$^{-13}$ & 1.55 & 216 & \citet{Reference509}\\ 
\ce{C2H4 + ^3O -> H2CO + ^3CH2} & 8.3$\times$10$^{-12}$ & 0 & 754 & \citet{Westenberg_DeHaas1969}\\ 
\ce{C2H4 + ^4N -> HCN + CH3} & 2.1$\times$10$^{-13}$ & 0 & 754 & \citet{Avramenko_Krasnenkov1964}\\ 
\ce{C2H4 + OH -> C2H3 + H2O -> C2H2 + H2O + H} & 1.7$\times$10$^{-13}$ & 2.75 & 2100 & \citet{Reference509}\\ 
\ce{C2H4 + CH3 -> C2H3 + CH4 -> C2H2 + CH4 + H} & 6.9$\times$10$^{-12}$ & 0 & 5600 & \citet{Reference451}\\ 
\ce{C2H4 + CN -> CH2CHCN + H} & 3.2$\times$10$^{-10}$ & 0 & 0 & \citet{Gannon_et_al2007}\\ 
\ce{C2H4 + CN -> HCN + C2H3 -> HCN + C2H2 + H} & 2.1$\times$10$^{-10}$ & 0 & 0 & \citet{Sayah_et_al1988}\\ 
\ce{C2H4 + ^2N -> CH3CN + H} & 2.2$\times$10$^{-10}$ & 0 & 500 & \citet{Hebrard_et_al2009},\\ 
 & & & & \citet{Balucani_et_al2012}\\
\ce{C2H6 + ^3CH2 -> C2H5 + CH3 -> C2H4 + CH3 + H} & 1.1$\times$10$^{-11}$ & 0 & 3980 & \citet{Reference568}\\ 
\ce{C2H6 + ^3O -> C2H5  + OH -> C2H4 + OH + H} & 8.6$\times$10$^{-12}$ & 1.5 & 2920 & \citet{Reference451}\\ 
\ce{C2H6 + H -> C2H5 + H2 -> C2H4 + H2 + H} & 1.2$\times$10$^{-11}$ & 1.5 & 3730 & \citet{Reference451}\\ 
\ce{C2H6 + OH -> C2H5 + H2O -> C2H4 + H2O + H} & 1.1$\times$10$^{-12}$ & 2.0 & 435 & \citet{Reference451}\\ 
\ce{C2H6 + CH -> C2H4 + CH3} & 1.3$\times$10$^{-10}$ & 0 & 0 & \citet{Galland_et_al2003}\\ 
\ce{C2H6 + CH -> CH3CHCH2 + H} & 3.0$\times$10$^{-11}$ & 0 & 0 & \citet{Galland_et_al2003}\\ 
\ce{C2H6 + CH3 -> CH4 + C2H5 -> CH4 + C2H4 + H} & 1.8$\times$10$^{-16}$ & 6.0 & 3040 & \citet{Reference451}\\ 
\ce{C2H6 + C2H -> C2H2 + C2H5 -> C2H2 + C2H4 + H} & 6.0$\times$10$^{-12}$ & 0 & 0 & \citet{Reference451}\\ 
\ce{C2H6 + CN -> HCN + C2H5 -> HCN + C2H4 + H} & 3.5$\times$10$^{-12}$ & 2.16 & 624 & \citet{Reference451}\\ 
\ce{C2H6 + ^1O -> C2H5 + OH -> C2H4 + OH + H} & 6.3$\times$10$^{-10}$ & 0 & 0 & \citet{Matsumi_et_al1993}\\ 
\ce{C2H6 + ^1O -> C2H6 + ^3O} & 7.3$\times$10$^{-10}$ & 0 & 0 & \citet{Fletcher_Husain1976}\\ 
\ce{CH3CHCH2 + CN -> CH3CN + C2H3 -> CH3CN + C2H2 + H} & 1.7$\times$10$^{-10}$ & 0 & -51 & \citet{Hebrard_et_al2009}\\ 
\ce{CH3CHCH2 + C2H -> CH2CHCH2 + C2H2} & 6.0$\times$10$^{-12}$ & 0 & 0 & \citet{Tsang1991}\\ 
\ce{CH3OH + ^4N -> CH3 + HNO} & 4.0$\times$10$^{-10}$ & 0 & 4330 & \citet{Roscoe_Roscoe1973}\\ 
\ce{CH3OH + OH -> H2CO + H2O + H} & 1.1$\times$10$^{-12}$ & 1.44 & 56 & \citet{Srinivasan_et_al2007}\\
\ce{CH3CHO + ^4N -> HCN + H2 + HCO} & 1.0$\times$10$^{-14}$ & 0 & 0 & \citet{Lambert_et_al1968}\\
\ce{CH3CHO + H -> CO + H2 + CH3} & 5.0$\times$10$^{-13}$ & 2.75 & 486 & \citet{Sivaramakrishnan_et_al2010}\\
\ce{CH3CHO + H -> CH4 + HCO} & 8.8$\times$10$^{-14}$ & 0 & 0 & \citet{Lambert_et_al1967}\\
 \ce{NH2 + ^3O -> H + HNO} & 7.5$\times$10$^{-11}$ & 0 & 0 & \citet{Reference1053}\\
  \ce{NH2 + ^3O -> OH + NH} & 1.2$\times$10$^{-11}$ & 0 & 0 & \citet{Reference1053}\\
    \ce{NH2 + ^3O -> H2 + NO} & 8.3$\times$10$^{-12}$ & 0 & 0 & \citet{Reference1053}\\
\ce{NH2 + NO -> N2 + H2O} & 5.9$\times$10$^{-11}$ & -2.37 & 437 & \citet{Song_et_al2001}\\
    \ce{NH2 + OH -> H2O + NH} & 7.7$\times$10$^{-13}$ & 1.5 & 230 & \citet{Reference1053}\\
    \ce{NH2 + OH -> NH3 + ^3O} & 5.0$\times$10$^{-15}$ & 2.6 & 870 & \citet{Reference1053}\\
\ce{NH2 + HO2 -> H2O + HNO} & 6.1$\times$10$^{-16}$ & 0.55 & 265 & \citet{Sumathi_Peyerimhoff1996}\\
\ce{NH2 + HO2 -> NH3 + O2} & 1.9$\times$10$^{-16}$ & 1.55 & 1020 & \citet{Sumathi_Peyerimhoff1996}\\
\ce{NH2 + H2 -> NH3 + H} & 2.1$\times$10$^{-12}$ & 0 & 4281 & \citet{Demissy_Lesclaux1980}\\
\ce{NH2 + C2H6 -> NH3 + C2H5 -> NH3 + C2H4 + H} & 6.1$\times$10$^{-13}$ & 0 & 3600 & \citet{Demissy_Lesclaux1980}\\
\ce{NH2 + CH4 -> NH3 + CH3} & 7.8$\times$10$^{-12}$ & 0 & 4680 & \citet{Moller_Wagner1984}\\
\ce{NH3 + ^3O -> NH2 + OH} & 1.6$\times$10$^{-11}$ & 0 & 3670 & \citet{Reference451}\\
\ce{NH3 + OH -> NH2 + H2O} & 3.5$\times$10$^{-12}$ & 0 & 925 & \citet{Atkinson_et_al2004}\\
\ce{NH3 + CN -> HCN + NH2} & 1.5$\times$10$^{-11}$ & 0 & 181 & \citet{Sims_Smith1988}\\
\ce{HNO + H -> H2 + NO} & 3.0$\times$10$^{-11}$ & 0 & 500 & \citet{Tsang_Herron1991}\\
\ce{HNO + ^3O -> OH + NO} & 3.8$\times$10$^{-11}$ & 0 & 0 & \citet{Inomata_Washida1999}\\
\ce{HNO + OH -> H2O + NO} & 8.0$\times$10$^{-11}$ & 0 & 500 & \citet{Tsang_Herron1991}\\
\ce{HNO + CN -> HCN + NO} & 3.0$\times$10$^{-11}$ & 0 & 0 & \citet{Reference509}\\
\ce{HNO + HCO -> H2CO + NO} & 1.0$\times$10$^{-12}$ & 0 & 1000 & \citet{Tsang_Herron1991}\\
\hline
\end{longtable*}

\begin{figure*}[!hbtp]
\centering
\includegraphics[width=\linewidth]{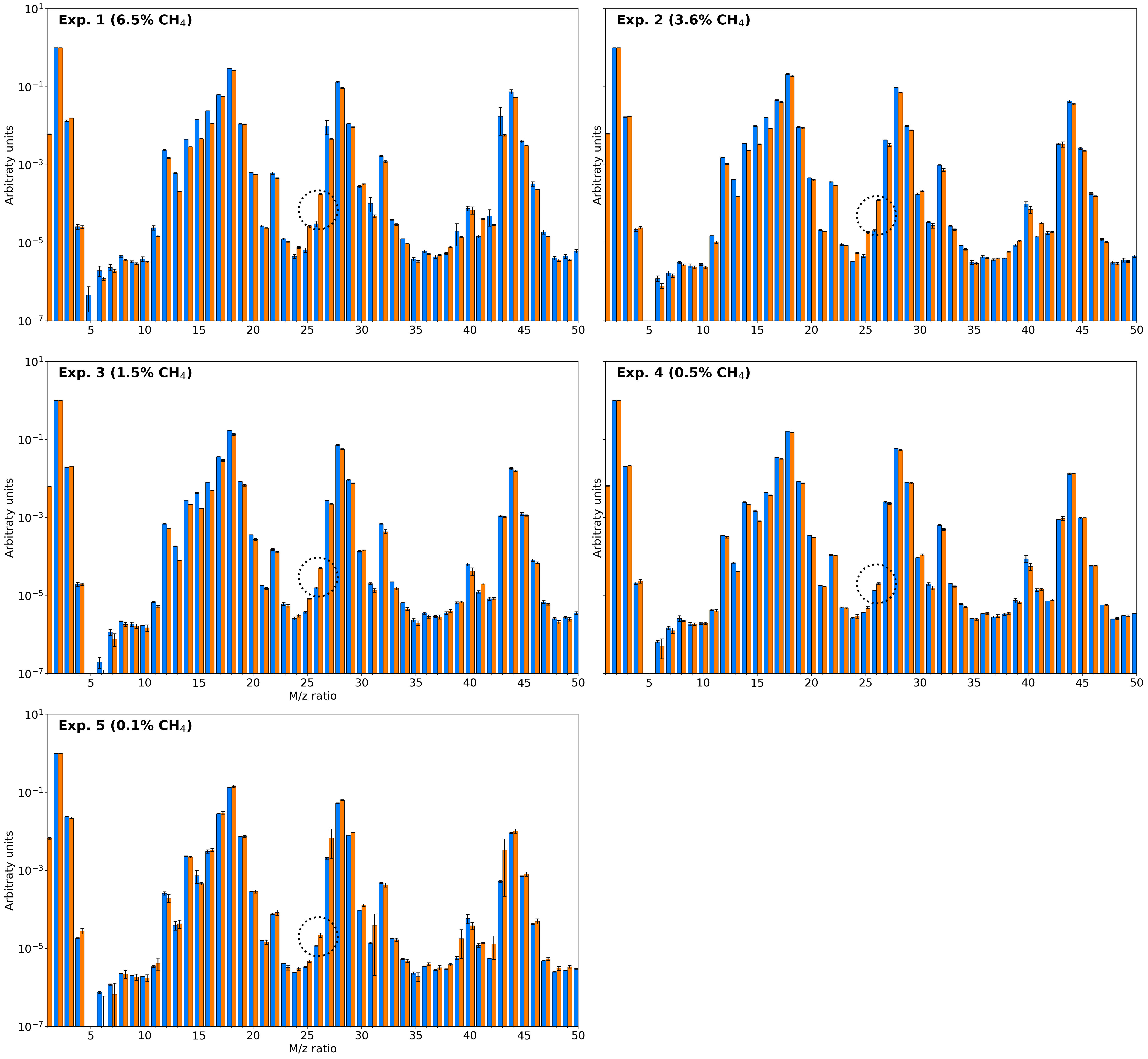}
\caption{Histogram mass spectra for Run 2 of Experiments 1--5. The 26 peak is circled in each plot to direct the reader's eye to the location where HCN can be detected in our experiments. \label{histograms2}}
\end{figure*}

In Figure~\ref{histograms2}, we plot Run 2 of Experiments 1--5. Run 1 data is displayed in the main text.

\begin{figure}[!hbtp]
\centering
\includegraphics[width=\linewidth]{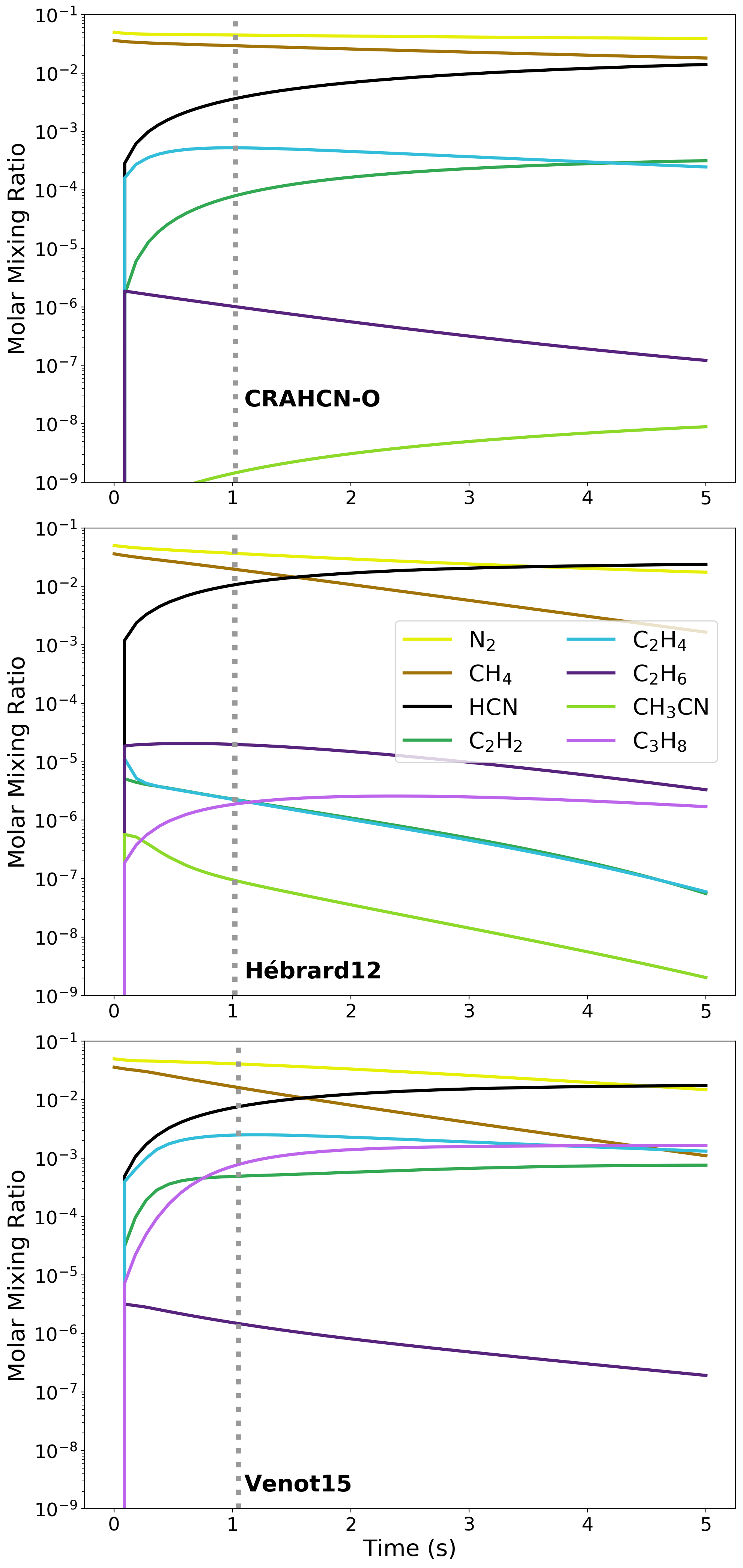}
\caption{Chemical kinetics simulations of Experiment 2 using three different chemical networks and the Maxwellian EEDF. The grey vertical dotted line represents the time step closest to $t$ = 1 second that is used to compare molecule concentrations across simulations. \label{Kintecus_Compare_Max}}
\end{figure}

In Figure~\ref{Kintecus_Compare_Max}, we investigate the chemical kinetics models of Experiment 2 using a Maxwellian EEDF, for the three chemical networks in this study. These results do not greatly differ from the Druyvestyn EEDF results in Figure 5 in the main text.

\bibliography{Bibliography}
\bibliographystyle{achemso}

\end{document}